\documentstyle[12pt,epsf,draft]{article}
\makeatletter
\def\vereq#1#2{\lower3pt\vbox{\baselineskip1.5pt \lineskip1.5pt
\ialign{$\m@th#1\hfill##\hfil$\crcr#2\crcr\gets\crcr}}}
\def\leriw{\mathrel{\mathpalette\vereq\to}}
\def\eqalign#1{\null\vcenter{\def\\{\cr}\openup\jot\m@th
  \ialign{\strut$\displaystyle{##}$\hfil&$\displaystyle{{}##}$\hfil
      \crcr#1\crcr}}\,}
\makeatother
\author{S.~S.~Sannikov\\
\it Physico-Technical Institute\\
\it 1 Academichna St., 310108 Kharkov, {\bf UKRAINE}\\[0.5 cm]
A.~A.~Stanislavsky\\
\it Institute of Radio Astronomy\\
\it of the Ukrainian National Academy of Sciences\\
\it 4 Chervonopraporna St., 310002 Kharkov, {\bf UKRAINE}\\
\it E-mail: stepkin@ira.kharkov.ua}
\title{NON-FOCK REPRESENTATIONS OF HEISENBERG ALGEBRAS}
\begin{document}
\large\tolerance8000\hbadness10000\emergencystretch3mm
\maketitle
\newtheorem{aff}{Statement}
\def\bbar #1{\bar {#1\vphantom{\bar #1}}}
\begin{abstract}
      The purpose of this paper is to present the mathematical
techniques of a new quantum scheme using a dual pair of reflexive
topological vector spaces in terms of the non-Hermitian form.
The scheme is shown to be a generalization of the well-known
unitary quantum theory and to describe jointly quantum objects and
physical vacuum.
\end{abstract}
{\qquad\small PACS number(s): 02.20.-a, 03.65.Fd, 03.70.+k}

\section*{I. Introduction}
      As is shown in \cite{1}, there is strong theoretical support
for the idea that under extremal conditions (at supersmall distances
or superhigh concentrations of energy) one has to take a
non-standard dynamical system (called relativistic bi-Hamiltonian one
described by the Heisenberg algebra $h^{(*)}_8$) into consideration.
In the case the Fock representation of the algebra $h^{(*)}_8$ and
unitary representation (generated by it) of its group of automorphisms
$Sp^{(*)}(4,\bf C)$ (a dynamical group of system) was found to be
incompatible to a condition of integrability of Hamiltonian flows
$p_\mu$ and $\dot p_\mu$ in a representation space, the Hilbert
space $H$. Hence, it is neccessary to use other representations
together with a more general scheme (called by the non-unitary
one for brevity) of functional analysis based the pair of
topological vector spaces $(\bf\dot F,F)$ dual with respect to
some non-Hermitian form $\langle\cdot,\cdot\rangle$. The theory
of such representations develops here.

      As far as we know the non-unitary quantum theory took its beginning
yet in the $60^{\mbox {\scriptsize st}}$ years when infinite dimensional
non-unitary representations of the rotation group and the
Lorentz group called later by semispinor ones were considered in \cite{2a}.
They result from the Dirac operation applied to Grassmann spinors.
A large part of these results (not on amount of publications, but
on volume of information) was presented in difficultly accessible
editions or as preprints.

      The mathematical apparatus of the new quantum theory is developed
here: non-Fock representations of those Heisenberg algebras which are
of interest from a physical standpoint are effectively built, their
connection with extended Fock representations including some additional
variables is established. The general purpose of this paper is to prove
the theorem formulating here in a conditional form so:
\begin{center}
\it The Non-unitary Quantum Theory = The Unitary\\
Quantum Theory $+$ hidden parameters . $(\ast)$
\end{center}

      If this theorem to read from the right on the left, it will
be possible to tell that an introduction of hidden (or additional)
variables in the usual (unitary) quantum theory makes it by non-unitary one,
i.\ e.\, as well as von Neuman \cite{3} assumed, this results in a
radical reorganization of theory.

      The paper is organized as follows. After a briefly review in
section~II of the Fock representation of Heisenberg algebras
describing the standard oscillator the spectrum of which is isomorphic
to the standard model of natural number series, we move directly in
section~III to the non-Fock representation and the non-standard oscillator
having an infinite number of states with negative numbers of occupation
and not having any ground state. The main technical obstacle seems to
be a cycling of creation and annihilation operators overcomed by means
of a decycling operation. In sections~IV-VI, we compare Fock and
non-Fock representations in detail. Finally, in sections~VII-VIII
physical consequences are discussed.

\section*{II. Fock representation of the algebra $h_2$ and
a representation of the algebra $sl(2,\bf C)$ associated with it}
\label{kd2}
      1) We start our consideration of Heisenberg algebra representations
connected in a certain way with Fock representations underlying the
Heisenberg-Schr$\rm\ddot o$dinger unitary quantum theory from a brief
reminder about the latters.

      The Heisenberg algebra $h_{2n}$ is normally written in term of
$2n$ generators $q_k, p_k (k=1,\ldots, n)$ by commutation relations
($h_{2n}$ are isomorphic to the nilpotent Lie algebra $n_{3n}$):
\begin{equation}
[q_j,p_k]=i\delta_{jk},\quad
[q_j,q_k]=[p_j,p_k] =0\,.\label{eq1}
\end{equation}
It is convenient to pass to Fock operators $a^{\rm a}_k$ where
\begin{equation}
a^1_k={1\over\sqrt{2}}(q_k+ip_k)\,,\quad
a^2_k={1\over\sqrt {2}}(q_k-ip_k)\label{eq2}
\end{equation}
obeyed commutation relations
\begin{equation}
[a^{\rm a}_k,a^{\rm a'}_{k'}]=\delta_{kk'}\,
\varepsilon^{\rm aa'},\quad{\rm a,a'}=1,2;\,k,k'=1,\ldots,n
\label{eq3}
\end{equation}
(here $\delta_{kk'}$ is the Cronecker symbol,
$\varepsilon^{\rm aa'}=\pmatrix{0&1\cr -1&0}$ is the
Levi-Civit$\rm\grave a$ symbol). The enveloping algebra $U[h_{2n}]$ represents
an infinite dimensional Lie algebra which contains a
finite dimensional Lie subalgebra denoted by
$l_{n(2n+3)+1}$ and called as a subalgebra of small oscillations,
structure of which is described by the formula (the Levi-Mal'tzev
decomposition)
\begin{equation}
l_{n(2n+3)+1}=(\sigma_{n(2n+1)}\oplus 1)\,+\!)\,h_{2n}
\label{eq4}
\end{equation}
where $\sigma_{n(2n+1)}\oplus 1=h_{2n}^2$, and $\sigma_{n(2n+1)}$
are isomorphic to the Lie algebra $sp(n,\bf C)$ ($h_{2n}$ is
understood to be the complex algebra $h_{2n}(\bf C)$). Generators  of
the algebra $\sigma_{n(2n+1)}$ are bilinear forms
$a^{\rm a}_k a^{\rm a'}_{k'}$.

      Putting the group $Sp(n,\bf C)$ as a group of internal
automorphisms of the algebra $h_{2n}(\bf C)$ , we shall formally write
\begin{equation}
T(v)a^{\rm a}_kT^{-1}(v)=v^{\rm aa'}_{kk'}a^{\rm a'}_{k'}
\label{eq5}
\end{equation}
where $v\in Sp(n,\bf C)$, and $T(v)$ is a ``spinor'' (infinite dimensional)
representation of the group $Sp(n,\bf C)$. By definition, Lie
brackets (\ref{eq3}) is invariant under transformations (\ref{eq5}).

      The Fock representation of the algebra $h_{2n}$ is built, as
is known \cite{4}, in the space
\begin{equation}
{\cal F}_F=\overline{U[a^2_k]}^{\,\tau}
\label{eq6}
\end{equation}
where $U[a^2_k]$ is the maximal commutative subalgebra in $U[h_{2n}]$
($a^2_k$ are coordinates on the Lagrangian plane), $\tau$ is a
topology. In the Fock realization we have $a^2_k=z_k\in\bf C$
and $a^1_k=\frac{\partial}{\partial z_k}$. Thus a Hermitian inner
product is defined on ${\cal F}_F$ as
\begin{equation}
(f,g)=\int_{{\bf C}^n}\overline{f(z)}\,g(z)\,d\mu(z),
\label{eq7}
\end{equation}
where $f,g\in{\cal F}_F(z=(z_1,\ldots,z_n)\in{\bf C}^n)$, with the
measure
\begin{equation}
d\mu(z)=\left(\frac{i}{2\pi}\right)^n\prod^n_{k=1}
e^{-\bbar z_kz_k}\,dz_k\wedge d\bbar z_k
\label{eq8}
\end{equation}
($\wedge$ is the external Cartan multiplication). In the
Heisenberg-Schr$\rm\ddot o$dinger quantum theory  the Hilbert
topology and the Hilbert space ${\cal F}_F={\cal H}$ is considered.
An additional symmetry of operators $a^{\rm a}_k$ or
$a^1_k=(a^2_k)^+$, where + is the conjugation for the
form (\ref{eq7}), connects with the inner product (\ref{eq7}):
\begin{equation}
(a^2_kf,g)=(f,a^1_kg).
\label{eq9}
\end{equation}
Therefore the Fock representation deals with the certain
real form of the algebra $h_{2n}(\bf C)$ denoted by
$h_{2n}(\bf R)$ having $Sp(n,\bf R)$ as a group of automorphisms.
As is known \cite{4}, the unitary two-valued
representation $v\to T(v)$ of the group $Sp(n,{\bf R})\ni v$
is realized in $\cal H$. The Fock representation is known also to
be unitary equivalent to the Schr$\rm\ddot o$dinebra representation
(\ref{eq1}) (the Stone-von Neuman theorem about uniqueness of unitary
representations of the Heisenberg group $H_{2n}$).

      We will be interested in other real forms of the algebra
$h_{2n}(\bf C)$ and the group $Sp(n,\bf C)$ (in particular, the
compact group $USp(n,{\bf C})=Sp(n)$) and their representations
which have completely another nature, as the operators (adequate
them) $T(v)$, first, are unbounded on the representation space
$\cal F$ (its topology will be determined further), and, secondly,
are multiple-valued on the group.

      The Fock representation of the algebra $h_{2n}(\bf C)$
proves to suffer by a serious lack: in some sense it is
inconsistently (this inconsistency is eliminated further),
and to see this, it is enough to consider the case $n=1$.

      2) We shall remind that the Dirac operation (see \cite{1})
applied to the Pauli bundle $E=({\bf A_{3,1}},S_2,\stackrel{-}{L})$
where a structural group $\stackrel{-}{L}$ of a fiber $S_2$ is the
group $SL(2,{\bf C})\approx Sp(1,{\bf C})$ results us in the
algebra $h_{2}(\bf C)$.

      As is known, the Fock representation of the algebra $h_{2}$
is built in the space
\begin{equation}
{\cal F}_F=\bigoplus^\infty_{m=0}f_m
\label{eq10}
\end{equation}
(here $\oplus$ is understood to be an orthogonal sum ) where
$\lbrace f_m\rbrace$ is a canonical basis on which
the operators $a^\alpha$ act under the law
\medskip
\begin{equation}
\def\maprile{\smash{\mathop{\leriw}\limits_{a^1}^{\ a^2}}}
0\enspace\smash{\mathop{\gets}\limits_{a^1}}\enspace f_0\enspace
\maprile\enspace f_1\enspace\maprile\enspace\cdots\enspace f_m\enspace
\maprile\enspace f_{m+1}\enspace\cdots
\label{eq11}
\bigskip
\end{equation}
i.\ e.\ $a^2 f_m\sim f_{m+1}$, and $a^1f_m\sim f_{m-1}$ ($f_0$
refers to as a ground state \cite{f1}:
$a^1f_0=0$). In the $z$-realization we have
\begin{equation}
a^\alpha={d/dz\choose z},\quad f_m=\frac{z^m}{\sqrt{m!}}\,,
\quad m=0,1,2,\ldots
\label{eq12}
\end{equation}

      Under the action of bilinear forms $a^\alpha a^\beta$
the space ${\cal F}_F$ breaks up in the orthogonal sum of
two subspaces of even and odd (rather $z\to-z$) functions:
${\cal F}_F={\cal F}^{(+)}\oplus{\cal F}^{(-)}$.

\begin{aff} Infinite dimensional irreducible representations
(adequate to spins $-\frac{1}{4}$ and $-\frac{3}{4}$) of the
Lie algebra $usp(1,{\bf C})\approx su(2)$ are realized in
${\cal F}^{(+)}$ and ${\cal F}^{(-)}$, are given by the operators
$\vec L=\frac{1}{4}a^\alpha\vec\sigma_\alpha^\beta a_\beta$
where $a_\beta=\varepsilon_{\beta\gamma}a^\gamma$
($\varepsilon_{\beta\gamma}=\pmatrix{0&1\cr -1&0}$, $\vec \sigma$
are the Pauli matrices) and are denoted by $D^+(-\frac{1}{4})$
and $D^+(-\frac{3}{4})$ \cite{2b}.
\end{aff}

      In the $z$-realization we have ($L_\pm=L_1\pm iL_2$)
\begin{equation}
L_3=\frac{1}{2}z\frac{d}{dz}+\frac{1}{4}\,,\quad
L_+=\frac{1}{2}z^2\,,\quad L_-=-\frac{1}{2}\frac{d^2}{dz^2}
\label{eq13}
\end{equation}
such that the Casimir operator is
$\vec L^2\equiv -\frac{3}{16}=\lambda (\lambda +1)$, whence and
it follows that $\lambda=-\frac{1}{4},-\frac{3}{4}$.

       $D^+(\lambda)$ is understood to be a representation of th
algebra $su(2)$ with the lowest Cartan vector (see \cite{2c} where such
representations refer to as semispinor ones) in a general case of
any spin $\lambda$. As $sl(2,{\bf C})=su^c(2)$, the
semispinor representation $D^+(\lambda)$ of the algebra $su(2)$
extends obviously up to the representation $(\lambda, 0)^+$
of the algebra $sl(2,{\bf C})$ \cite{2d}.

      Next, according to be told above let us introduce the denotions
\begin{displaymath}
{\cal F}^{(+)}={\cal F}^{(+)}_{-\frac {1}{4}}\,,\quad
{\cal F}^{(-)}={\cal F}^{(-)}_{-\frac {3}{4}}\,.
\end{displaymath}
By definition,
\begin{equation}
a^\alpha:\quad{\cal F}^{(-)}_{-\frac{3}{4}}
\leftarrow{\cal F}^{(+)}_{-\frac{1}{4}} \,,
\label{eq14}
\end{equation}
so the operators $a^\alpha$, changing a parity of space,
lower a weight of representation (spin) on $\frac{1}{2}:\quad
-\frac{1}{4}\to-\frac{1}{4}-\frac{1}{2}=-\frac{3}{4}$.
On the other hand, in the Fock representation
\begin{equation}
a^\alpha:\quad{\cal F}^{(-)}_{-\frac{3}{4}}
\rightarrow{\cal F}^{(+)}_{-\frac{1}{4}} \,,
\label{eq15}
\end{equation}
i.\ e.\ the same operators raise the weight of representation on
$\frac{1}{2}:\quad -\frac{3}{4}\to-\frac{3}{4}+\frac{1}{2}=
-\frac{1}{4}$. The phenomenon described by the formulae (\ref{eq14}),
(\ref{eq15}) refers us to as a cycling of the operators
$a^\alpha$, and here is this cycle:
\medskip
\begin{displaymath}
{\cal F}^{(-)}_{-\frac{3}{4}}\quad
\smash{\mathop{\leriw}\limits_{a^\alpha}^{\ a^\alpha}}
\quad{\cal F}^{(+)}_{-\frac{1}{4}}\,.
\bigskip
\end{displaymath}

      The given phenomenon is connected with a weight of representation
(spin, as the speech goes actually about representations of the
rotation group $SO(3)\sim SU(2)\approx Sp(1)$) taking on values
in the numerical field $\bf Z_2$ of the simple characteristic 2 \cite{2e}.
Indeed, it is possible to present the pair of numbers $-\frac{1}{4}$
and $-\frac{3}{4}$ as $\lambda=-\frac{1}{4}-\frac{p}{2}$
where $p\in{\bf Z_2}=\lbrace 0,1\rbrace$. Thus we can write
\begin{equation}
{\cal F}_F=\bigoplus_{p\in{\bf Z_2}}{\cal F}^{((-1)^p)}_{-\frac{1}
{4}-\frac{p}{2}}\,.
\label{eq16}
\end{equation}
It is an inconsistency of the Fock representation of the algebra $h_2$
from the points of view of the theory of spin that consists that in
this representation a spin takes on values in the field $\bf Z_2$,
while on physical reasons it must take on values in the field
of zero characteristic. This remark is of extremely importance for
whole subsequent consideration.

      The decycling operation formulated further is connected with
a passage from the ring $\bf Z_2$ to the standard ring of zero
characteristic $\bf Z$, which is connected with $\bf Z_2$ by the formula
${\bf Z_2}={\bf Z}/\!\!\bmod 2$. From the point of view of this formula
$\bf Z$ is a universal covering for $\bf Z_2$.

\section*{III. Decycling operation.
Non-Fock representation of the algebra $h_4$}\label{kd3}
      1) At first we shall describe the non-Fock representation of
the algebra $h_2$ \cite{2e}.

      As is known, the Fock representation (denoted as $T_0(h_2)$)
of the algebra $h_2$ plays an important role in
the Heisenberg-Schr$\rm\ddot o$dinger quantum (unitary) theory. It
is completely characterized by its ground state $\vert 0\rangle$,
called sometimes the mathematical vacuum, or the spectrum of the
operator $\stackrel{\wedge}{N}=a_2a_1=a^+_1a_1$, which is isomorphic to
the standard model of natural number series
${\bf Z}_+=\lbrace 0,1,2,\ldots\rbrace$. The standard oscillator
(it has a ground state) is described by this representation.

      Another non-standard (or non-Fock) representation
denoted by $T_1(h_2)$ is possible to underlie another quantum
theory --- the non-unitary one. A refusal from an additional
condition of Hermitian symmetry $a_2=a^+_1$, and, hence, from
a condition of the positive definite spectrum for the operator
$\stackrel{\wedge}{N}=a_2a_1$ results in the representation
$T_1(h_2)$. In these more general conditions a spectrum of the
operator $\stackrel{\wedge}{N}$, remaining equivalent, becomes
unlimited in both parties, and in general it is complex:
$Sp\,\stackrel{\wedge}{N}=\lbrace\varepsilon +p\rbrace_{p\in\bf Z}$.
Weight vectors $\vert p\rangle$ of such (non-standard) oscillator
satisfy the equation
$\stackrel{\wedge}{N}\vert p\rangle=(\varepsilon +p)\vert u\rangle$,
and the spectrum of the operator $\stackrel{\wedge}{N}$ is isomorphic to
a non-standard model of natural number series
${\bf Z}=\lbrace \ldots,-2,-1,0,1,2,\ldots\rbrace$ ($\bf Z$
satisfies all the axiomas of arithmetics, excepting the axioma of choice).

       $\varepsilon\in{\bf C}/{\bf Z}$ associates with
inequivalent representations such that for $\varepsilon\ne 0$
a representation is irreducible (for $\varepsilon=0$ it is
incompletely reducible, as there is an invariant subspace strained
on vectors $\vert 0\rangle,\vert 1\rangle,\vert 2\rangle,\ldots$). As
is seen, $h_2$ has many non-standard representations, but all of them
describe the same object --- the non-standard oscillator.

      The non-standard oscillator has an infinite set of states
with negative numbers of occupation:
$\vert -1\rangle,\vert -2\rangle,\ldots$ (which we call
by states of physical vacuum), and, hence, does not have
a ground state ( the first element is not present). ${\bf Z}$
is said to be ordered as a set, while ${\bf Z_+}$
is completely ordered. However according to the Zermelo theorem any
set can be completely ordered. Our basic result --- the theorem
$(\ast)$ about existence of additional (hidden) parameters in
the non-unitary theory is predetermined by that theorem
(see section~V).

      2) We shall return to the Fock representation of the algebra $h_2$.
Taking into account the physical nature of spin (see above),
it is necessary to pass from the space ${\cal F}_F$ to the space \cite{2f}
\begin{equation}
{\bf F}_z=\bigoplus_{p\in
{\bf Z}}{\cal F}^{((-1)^p)}_{-\frac{1}{4}+\frac{p}{2}}
\label{eq17}
\end{equation}
which is a sort of covering for the pair of spaces
(\ref{eq16}) (as ${\bf Z}$ is a covering for
${\bf Z_2}$). Here $\oplus$ is understood to be
a direct sum of spaces, as an opportunity of the usual
sum are already exhausted (we shall notice that in the theory of
representations of the group
$SO(2,1)\sim SU(1,1)\approx Sp(1,{\bf R})$ a weight of
representation is not any spin, and consequently the expansion
${\cal F}_F\subset{\bf F}_z$ is not obligatory in this case).

      On the space ${\bf F}_z$ (\ref{eq17}) the operators $a^\alpha$
act under the law
\begin{displaymath}
\def\mapleft{\smash{\mathop{\gets}\limits_a}}
\cdots\quad\mapleft\quad{\cal F}^{(+)}_{-\frac{5}{4}}\quad\mapleft
\quad{\cal F}^{(-)}_{-\frac{3}{4}}\quad\mapleft\quad
{\cal F}^{(+)}_{-\frac{1}{4}}\quad\mapleft\quad\cdots,
\bigskip
\end{displaymath}
i.\ e.\ they always only lower spin $\lambda$ on
$\frac{1}{2}$ ($\lambda\to\lambda-\frac{1}{2}$), so
$a^\alpha$: ${\cal F}^{((-1)^p)}_{-\frac{1}{4}+\frac{p}{2}}
\,\gets\,{\cal F}^{((-1)^{p-1})}_{-\frac{3}{4}+\frac{p}{2}}$.
The operators raising spin on $\frac{1}{2}$
($\lambda\to\lambda +\frac{1}{2}$) are denoted by
$b_\alpha$ and act on ${\bf F}_z$ under the law:
\medskip
\begin{displaymath}
\def\mapright{\smash{\mathop{\to}\limits^b}}
\cdots\quad\mapright\quad{\cal F}^{(-)}_{-\frac{3}{4}}\quad\mapright
\quad{\cal F}^{(+)}_{-\frac{1}{4}}\quad\mapright\quad
{\cal F}^{(-)}_{\frac{1}{4}}\quad\mapright\quad\cdots,
\end{displaymath}
i.\ e.\ $b_\alpha$: ${\cal F}^{((-1)^{p-1})}_{-\frac{3}{4}
+\frac{p}{2}}\,\gets\,{\cal F}^{((-1)^p)}_{-\frac{1}{4}+\frac{p}{2}}$.
The operators $b_\alpha$ were found in \cite{2g}, and in
the $z$-realization they are written down as
\begin{equation}
b_\alpha=\,^{(p)}b_\alpha=\frac{1}{2}\left(z,
-\frac{d}{dz}+\frac{2p}{z}\right)\,,\quad p\in\bf Z
\label{eq18}
\end{equation}
With $a^\alpha$ they are connected by the relation (it is checked
directly)
\begin{equation}
^{(p)}\!b_\alpha=z^{2p} a^\beta z^{-2p}\varepsilon_{\beta\alpha}\,.
\label{eq19}
\end{equation}
As is seen, it is the phenomenon of cycling that is connected
with a degeneracy at $p=0$:
$^{(0)}\! b_\alpha=\frac{1}{2}a^\beta\varepsilon_{\beta\alpha}$.

\begin{aff} On the space ${\bf F}_z$
the operators $b_\alpha$ satisfy relations
\begin{displaymath}
^{(p+1)}\!b_\alpha\,\,^{(p)}\!b_{\alpha'}\,\,-
\,\,^{(p+1)}\!b_{\alpha'} \,\,^{(p)}\!b_\alpha=
\varepsilon_{\alpha\alpha'}\,,
\end {displaymath}
and, hence, set some representation of the algebra $h_2$
on ${\bf F}_z$. This representation (as well as the
representation given by the operator $a^\alpha$) is incompletely
reducible, as in ${\bf F}_z$ there is a flag
$0\subset\cdots\subset{\bf F}_z^N\subset
{\bf F}_z^{N+1}\subset\cdots\subset{\bf F}_z$
from $h_2$-invariant subspaces
\begin{displaymath}
{\bf F}_z^N=\bigoplus_{p=-\infty}^N
{\cal F}^{((-1)^p)}_{-\frac{1}{4}+\frac{p}{2}}\,.
\end{displaymath}
\end{aff}

      It is important to notice that the operators $a$ and $b$ do not
commute among themselves. Further, these operators will be transformed in
the operators $\varphi$ and $\bbar\varphi$ following absolutly
to other commutation relations (\ref{eq20}).

      Now note that the space ${\bf F}_z$
contains subspaces of different parity having
some inconveniences. Moreover the $z$-realization possesses
a defect such that the operators $^{(p)}\!b_\alpha$ belong
not to the algebra $U[h_2]$, but the divisive ring
$K(h_2)=S^{-1}U[h_2]$ (entered in \cite{2g}) builded over $h_2$ on
the multiplicative subset $S=U[b_1]$. All this
forces, making use of a connection between spaces of
different parity ${\cal F}^{(-)}_\lambda=z{\cal F}^{(+)}_\lambda$,
to pass to spaces of one, namely, positive parity. As
\begin{displaymath}
a^\alpha{\cal F}^{(+)}_\lambda\subseteq
{\cal F}^{(-)}_{\lambda-\frac{1}{2}} =
z {\cal F}^{(+)}_{\lambda-\frac{1}{2}}
\end{displaymath}
and
\begin{displaymath}
b_\alpha{\cal F}^{(-)}_\lambda=
b_\alpha z{\cal F}^{(+)}_\lambda\subseteq
{\cal F}^{(-)}_{\lambda +\frac{1}{2}}\,,
\end{displaymath}
we shall have
\begin{displaymath}
\frac{1}{z}a^\alpha{\cal F}^{(+)}_\lambda\subseteq
{\cal F}^{(+)}_{\lambda-\frac{1}{2}}\,,\quad
b_\alpha z{\cal F}^{(+)}_\lambda\subseteq
{\cal F}^{(-)}_{\lambda +\frac{1}{2}}\,.
\end{displaymath}

{\bf Theorem }{\it\ The operators
\begin{displaymath}
\varphi=\frac{1}{z}a={z^{-1}d/dz\choose 1},\quad
^{(p)} \!\bbar\varphi=\frac {1} {2} \,\,^{(p)}\!b_\alpha\,z=
\frac{1}{2}\left(\,z^2\,,\,-z\frac{d}{dz}+2p+1\,\right)
\end{displaymath}
are densely defined on the space
\begin{displaymath}
{\bf F}_z^{(+)}=\bigoplus_{p\in{\bf Z}}
{\cal F}^{(+)}_{-\frac{1}{4}+\frac{p}{2}}\,,
\end{displaymath}
act on it under the law
\medskip
\medskip
\begin{displaymath}
\def\maprile#1{\smash{\mathop{\leriw}
\limits_{\varphi}^{^{\scriptstyle(#1)}\!\bbar\varphi}}}
\cdots\enspace\maprile{-1}\enspace {\cal F}^{(+)}_{-\frac{3}{4}}
\enspace \maprile 0\enspace {\cal F}^{(+)}_{-\frac{1}{4}}
\enspace\maprile 1\enspace {\cal F}^{(+)}_{\frac{1}{4}}
\enspace\maprile 2\enspace\cdots
\bigskip
\end{displaymath}
and obey commutation relations
\begin{equation}
\eqalign{
\varphi^\alpha\varphi^{\alpha'}-\,\,\varphi^{\alpha'}
\varphi^\alpha=\,\,^{(p+1)}\!\bbar\varphi_\alpha\,\,
^{(p)}\!\bbar\varphi_{\alpha'}\,\,-\,\,^{(p+1)}\!\bbar
\varphi_{\alpha'}\,\,^{(p)}\!\bbar\varphi_\alpha=
0,\\ \varphi^\alpha\,\,^{(p+1)}\!\bbar
\varphi_{\alpha'}\,\,-\,\,^{(p)}\!\bbar\varphi_{\alpha'}
\,\,\varphi^\alpha=\delta^\alpha_{\alpha'}.}
\label{eq20}
\end{equation}}

      We shall enter the operators $a_\alpha^{\rm a}$, defining them
by their restrictions on each of subspaces
${\cal F}^{(+)}_{-\frac{1}{4}+\frac{p}{2}}\,(p\in {\bf Z})$:
\begin{equation}
a^1_\alpha\vert_{{\cal F}^{(+)}_{-\frac{1}{4}+\frac{p}{2}}}=
\varphi^\alpha\,,\quad
a^2_\alpha\vert_{{\cal F}^{(+)}_{-\frac{1}{4}+\frac{p}{2}}}=
\,\,^{(p)}\!\bbar\varphi_\alpha\,.
\label{eq21}
\end{equation}
(the language of Hopf algebras could here be pertinent, however it is
too cumbersome \cite{2h}).

      From (\ref{eq20}) it follows that on ${\bf F}_z^{(+)}$
the operators $a_\alpha^{\rm a}$ satisfy commutation relations
\begin{equation}
[a_\alpha^{\rm a},a_{\alpha'}^{\rm a'}]=\delta_{\alpha\alpha'}\,
\varepsilon^{\rm aa'}\,,
\label{eq22}
\end{equation}
and by that set some representation of the Heisenberg algebra
$h_4$ on ${\bf F}_z^{(+)}$, which, as it will be
shown further, is not equivalent to the Fock representation of this
algebra.

      Let's assume $\frac{1}{2}z^2=\zeta$. Then the operators
$\varphi^\alpha,\,\,^{(p)}\!\bbar\varphi_\alpha$ will be written down in
the form
\begin{equation}
\varphi^\alpha={d/d\zeta \choose 1},\quad
^{(p)}\!\bbar\varphi_\alpha=\left(\zeta,\,
-\zeta\frac{d}{d\zeta}+p+\frac{1}{2}\right).
\label{eq23}
\end{equation}
In this realization the space of representation is written down
as
\begin{equation}
{\bf F}_\zeta=\bigoplus_{p\in {\bf Z}}
{\cal F}_{-\frac{1}{4}+\frac{p}{2}}
\label{eq24}
\end{equation}
where ${\cal F}_{-\frac{1}{4}+\frac{p}{2}}$ are classes of holomorphic
functions of a complex variable $\zeta$.

      Points $-\frac{1}{4}+\frac{p}{2}$ of a complex plane of
a spin variable $\lambda$ have a unpleasant
property that between two points $-\frac{1}{4}$ and $-\frac{3}{4}$
the cycling is still possible, and this phenomenon
to exclude for ever, we shall replace $-\frac{1}{4}$ by a point of the
general position $\lambda\ne-\frac{1}{4},0,\frac{1}{2}$,
assuming a spin value in the formulae (\ref{eq23}), (\ref{eq24}), is equal
$\lambda +\frac{p}{2}$. We write down these formulae as
\begin{equation}
\eqalign{
\varphi^\alpha={d/d\zeta\choose 1},\quad
^{(p)}\!\bbar\varphi_\alpha=\left(\zeta,\,
-\zeta\frac{d}{d\zeta}+2\lambda +p+1\right),\\
{\bf F}^{(\lambda)}_\zeta=\bigoplus_{p\in{\bf Z}}
{\cal F}_{\lambda+\frac{p}{2}}}
\label{eq25}
\end{equation}
((\ref{eq25}) pass to (\ref{eq23}) at $\lambda=-\frac{1}{4}$).
Thus we shall have
\begin{equation}
\varphi^\alpha:\,\,{\cal F}_{\lambda +\frac{p}{2}} \,\to
\,{\cal F}_{\lambda +\frac{p-1}{2}}\,;\quad
^{(p)}\!\bbar\varphi_\alpha:\,\,{\cal F}_{\lambda +\frac{p}{2}}
\,\to\,{\cal F}_{\lambda +\frac{p+1}{2}}\,.
\label{eq26}
\end{equation}

\section*{IV. Comparison of representations}\label{kd4}
\begin{aff}\label{San3}
A representation of algebra $h_4$ in the space ${\bf F}_\zeta$
is not equivalent to the Fock representation of this algebra, which is
realized in the space ${\cal F}_F$.
\end{aff}

      To see this, at first we shall give basic formulae of the
Fock representation of the algebra $h_4$.

      1) This representation is given by the operators
\begin{equation}
a^1_\alpha=\frac{\partial}{\partial z_\alpha}=\varphi^\alpha,
\quad a^2_\alpha=z_\alpha=\bbar\varphi_\alpha,\quad
\alpha=1,2;\quad z_\alpha\in{\bf C}
\label{eq27}
\end{equation}
(see section~II) acting on the Fock space
${\cal F}_F$ formed by functions of two complex
variables $z_1,\,z_2$. A canonical basis in ${\cal F}_F$
consists of own functions of the operator $N=\sum_{\alpha=1}^2
a^2_\alpha a^1_\alpha$~, which are written down as
\begin{displaymath}
f^{m_2}_{m_1}=\frac{z^{m_1}_1\,z^{m_2}_2}{\sqrt{m_1!m_2!}}\,,
\quad m_1,m_2=0,1,2,\ldots
\end{displaymath}
In this basis we have the formulae
\begin{equation}
\eqalign{
\varphi^1f^{m_2}_{m_1}=\sqrt{m_1}f^{m_2}_{m_1-1}\,,\quad
\bbar\varphi_1 f^{m_2}_{m_1}=\sqrt {m_1 +1} f^{m_2}_{m_1 + 1}\,,\\
\varphi^2f^{m_2}_{m_1}=\sqrt{m_2}f^{m_2 -1}_{m_1}\,,\quad
\bbar\varphi_2f^{m_2}_{m_1}=\sqrt{m_2 +1}f^{m_2 +1}_{m_1}}
\label{eq28}
\end{equation}
from which it follows that the Fock representation describes a pair of
standard oscillators having the same ground state $f^0_0$.
The representation is denoted by us as $T_0(h_4)$.

      An important characteristic of the representation of the algebra
$h_4$ is that which representation of the algebra $su(2)$ giving
by the operators $\vec L=\frac{1}{2}a^2\vec\sigma a^1$ is realized
in the representation space of the algebra $h_4$. In the Fock
representation the operators $\vec L$ is of the following form
\begin{displaymath}
L_3=\frac{1}{2}\left(z_1\frac{\partial}{\partial z_1}-
z_2\frac{\partial}{\partial z_2}\right),\quad
L_+=z_1\frac{\partial}{\partial z_2},\quad
L_-=z_2\frac{\partial}{\partial z_1}\,.
\end{displaymath}
In ${\cal F}_F$ they result in a representation of the algebra $su(2)$
in the form
\begin{equation}
\bigoplus_{p\in{\bf Z_+}}D(p/2)\,.
\label{eq29}
\end{equation}
Here $D(p/2)$ is a finite dimensional representation with spin
$p/2$, which is realized in the subspace of homogeneous
polinomials of a degree $p$ in the form $\sum_{k=0}^p a_k z_1^k z_2^{p-k}$.
The Casimir operator $\vec L^2=L_0(L_0+1)$, where $L_0=\frac{1}{2}
a^2a^1=\frac{1}{2}\left (z_1\frac{\partial}{\partial z_1}+
z_2\frac{\partial}{\partial z_2}\right)$, takes on values
$\frac{1}{2}p(\frac{1}{2}p+1)$ on such polinomials. Hence,
on ${\cal F}_F$ the representation $su(2)$ is completely reducible
to its finite dimensional representations.

      A representation of the wider, Lorentz algebra $sl(2,{\bf C})$
giving by the operators $\vec L,\,\vec N$, where $\vec N=\frac{i}{2}
(a^2\vec\sigma\varepsilon a^2+a^1\varepsilon\vec\sigma a^1)$,
is characterized by the following values of the Casimir operators:
$C=\vec L^2-\vec N^2=-\frac{3}{4},\,C'=\vec L\vec N=0$. On
${\cal F}_F$ these operators set the representation $[\frac{1}{2},
0]\oplus[0,\frac{1}{2}]$\,\cite{f2}
where $[\frac{1}{2},0],\,[0,\frac{1}{2}]$ are the well-known
infinite dimensional Majorana representations of the group $SL(2,{\bf C})$
\cite{5}.

      2) The representation constructed by us for algebra $h_4$ (\ref{eq25})
first differs from the Fock one by being given in the
space of functions of one complex variable $\zeta$ instead of
two. Moreover it differs by on the space
${\bf F}_\zeta^{(\lambda)}$ a representation (generated by it) of the algebra
$su(2)$ (giving by the operator $\vec L=\frac{1}{2}a^2\vec\sigma a^1$)
being built of infinite dimensional semispinor representations
$D^+(\lambda)$ of the algebra $su(2)$ under the following formula \cite{2e}:
\begin{equation}
\bigoplus_{p\in {\bf Z}}D^+(\lambda +p/2)\,.
\label{eq30}
\end{equation}
The representation $D^+(\lambda +p/2)$ is realized in the subspace
${\cal F}_{\lambda +\frac{p}{2}}$ and is given by the operators
$\vec L^{(\lambda +\frac{p}{2})}$ being a
restriction of the operators $\vec L$ on the subspace
${\cal F}_{\lambda +\frac{p}{2}}$: $\vec L^{(\lambda +\frac{p}{2})}=
\vec L\vert_{{\cal F}_{\lambda +\frac{p}{2}}}=\frac{1}{2}\,\,
^{(p)}\!\bbar\varphi\vec\sigma\varphi$ where $\varphi,\,\bbar\varphi$
are defined by the formulae (\ref{eq25}), (\ref{eq26}), whence it follows
that
\begin{equation}
L^{(\lambda +\frac{p}{2})}_3=\zeta\frac{d}{d\zeta}-(\lambda +\frac{p}{2})
,\quad L^{(\lambda + \frac {p}{2})}_+ =\zeta,\quad
L^{(\lambda +\frac {p}{2})}_-=-\zeta\frac{d^2}{d\zeta^2}+2(\lambda+
\frac{p}{2})\frac{d}{d\zeta}
\label{eq31}
\end{equation}
being $\left(\vec L^{(\lambda +\frac{p}{2})}\right)^2=
\lambda(\lambda +1)$. As is seen, we deal here with a non-Lie
realization (differential operators of the second order; compare
with the Fock realization, the formulae (\ref{eq27}) -- (\ref{eq29})).
The Cartan-Weyl basis in ${\cal F}_{\lambda +\frac{p}{2}}$ is formed
by functions
\begin{equation}
f^{(\lambda +\frac{p}{2})}_{m(\zeta)}=(i\zeta)^m/
\sqrt{m!\,\Gamma (m-2\lambda-p)}\,,\quad m=0,1,2,\ldots,
\quad p\in{\bf Z}
\label{eq32}
\end{equation}
where $\Gamma$ is the Euler function normalized by the condition
\begin{displaymath}
\langle f^{(\bar\lambda)}_m\,,f^{(\lambda)}_{m'}
\rangle_\lambda=(-1)^m\delta_{mm'}
\end{displaymath}
where
\begin{equation}
\langle f,g\rangle_\lambda=\int\overline{f(\zeta)}\,Ig(\zeta)\,
d\mu_\lambda (\zeta),
\label{eq33}
\end{equation}
$Ig(\zeta)=g(-\zeta)$, and
\begin{equation}
d\mu_\lambda(\zeta)=\frac{i}{\pi}\frac{1}{\vert\zeta
\vert^{2\lambda +1}}K_{2\lambda +1}(2\vert\zeta\vert)\,
d\zeta\wedge d\bar\zeta
\label{eq34}
\end{equation}
($K_a$ is the Macdonald function, this measure is a generalisation
of the Gauss measure on the case of any spin $\lambda$) is
an $su(2)$-invariant sesquilinear form on
${\cal F}_{\lambda +\frac{p}{2}}$ determined in \cite{2i}:
$\langle f^{(\bar\lambda)}\,,\vec L^{(\lambda)}g^{(\lambda)}
\rangle_\lambda=\langle\vec L^{(\bar\lambda)}f^{(\bar\lambda)}
\,,g^{(\lambda)}\rangle_\lambda$. Thus the operators
$\vec L^{(\bar\lambda)}$ answering a complex conjugate
spin $\bar\lambda$, set a conjugate representation
$D^+(\bar\lambda)$ in the dual space ${\cal F}_{\bar\lambda}$.
It is necessary to notice that the form (\ref{eq33}) is
not invariant under transformations $su^c(2)=su(2)+i\,su(2)$.

      On the whole space ${\bf F}_\zeta^{(\lambda)}$
the sesquilinear form is given by the sum
\begin{equation}
\langle f,g\rangle=\sum^\infty_{p=-\infty}
\langle f^{(\bar\lambda +\frac{p}{2})}\,,g^{(\lambda +\frac{p}{2})}
\rangle_{\lambda +\frac{p}{2}}
\label{eq35}
\end{equation}
where $g^{(\lambda +\frac{p}{2})}$ is a projection of $g$ in
the space ${\cal F}_{\lambda +\frac{p}{2}}$\,. Thus, it is
obvious that $\langle f,\vec Lg\rangle=\langle\vec Lf,g\rangle$.
Moreover, the form (\ref{eq35}) is invariant under the
wider algebra $sp(2,{\bf R})$, generators of which
are $I_{\mu\nu}=(\vec L,\vec N),\,\Gamma_\mu=(\vec\Gamma,
\Gamma_0)$ where $\vec N=\frac{i}{4}(a^1\varepsilon\vec\sigma a^1+
a^2\vec\sigma\varepsilon a^2),\,\vec\Gamma=\frac{1}{4}
(a^1\varepsilon\vec\sigma a^1-a^2\vec\sigma\varepsilon a^2),\,
\Gamma_0=L_0+\frac{1}{2}$, and $L_0=\frac{1}{2}a^2a^1=\frac{1}{2}N$
such that $\vec L^2=L_0(L_0+1)$. In the given representation the Casimir
operators have the same values $C=\vec L^2-\vec N^2=-\frac{3}{4}\,,
C'=\vec L\vec N=0\,,\Gamma_\mu^2=\vec\Gamma^2-\Gamma_0^2=\frac{1}{2}$,
as in the Fock representation, though the representations, as we
see, are different. The point is that values of the Casimir operators
do not define a representation of the algebra $sl(2,{\bf R})$ completely.
The spectrum of the operator $L_0$, which consists of points
$Sp\,L_0=\lbrace\lambda +\frac{p}{2}\rbrace_{p\in{\bf Z}}$
on the space ${\bf F}_\zeta^{(\lambda)}$, is important too, while on the
space ${\cal F}_F$ it is formed by points $Sp\,L_0=\lbrace\frac{p}{2}
\rbrace_{p\in{\bf Z_+}}$. The representation of the algebra $sl(2,{\bf C})$
on ${\bf F}_\zeta^{(\lambda)}$ is denoted as $[\frac{1}{2},0]^+\oplus
\,[0,\frac{1}{2}]^+$ where $[\frac{1}{2},0]^+$ is an infinite
dimensional ``tail'' of the Majorana representation $[\frac{1}{2},0]$
unequivalent to the Majorana representation. By a restriction of
$sl(2,{\bf C})\supset su(2)$ the given representation breaks up on
irreducible semispinor representations $su(2)$ under the formula
(\ref{eq30}).

      At last we mention one more remark. Let us denote
$f^\mu_m=f^{(\lambda +\frac{p}{2})}_m$ to put
$\mu=2\lambda +p-m$. In this basis the operators $a^{\rm a}_\alpha$
are given by the formulae \cite{2e}:
\begin{equation}
\eqalign{
a^1_1f^{\mu}_m=\sqrt{m}f^\mu_{m_1-1}\,,\quad
a^2_1f^{\mu}_m=\sqrt{m+1}f^\mu_{m+1}\,,\\
a^1_2f^{\mu}_m=\sqrt{\mu} f^{\mu -1}_m\,,\quad
a^2_2f^{\mu}_m=\sqrt{\mu +1}f^{\mu +1}_m\,,}
\label{eq36}
\end{equation}
from which it follows that the operators $(a^1_1\,,a^2_1)$
are related to the standard oscillator (there is a ground state, as
$m=0,1,2,\ldots$), while
$(a^1_2\,,a^2_2)$ apply to the non-standard oscillator
(a ground state is not present, as $p=0,\pm 1,\pm 2,\ldots$).
The representation $h_4$ is denoted by us as $T_1(h_4)$
and refers to as non-Fock one. A representation, in which both
oscillators are non-standard, is denoted by $T_2(h_4)$.

      Now it is possible to say definitely that
the representations $T_0(h_4)$ and $T_1(h_4)$ are not equivalent
(Statement \ref{San3}).

      In general the algebra $h_{2n}$ has $n+1$ representations
$T_k(h_{2n})$ (unequivalent among themselves) where
$k\,(0\leq k\leq n)$ is the number of non-standard oscillators.

      We shall give yet a precise definition of decycling for
the Fock representation $T_0(h_{2n})$. The maximal decycling is
understood to be a passage from the Fock representation
$T_0(h_{2n})$ of the algebra $h_{2n}$ to the non-Fock representation
$T_n(h_{4n})$ of the algebra $h_{4n}$ in twice of greater dimension.

      3) At the end of this section we shall stop on the description
of semispinor representations of the algebra $su(2)$, its complex
expansions $su^c(2)$ and also the group $SU^c(2)$ as a whole. As
we already noticed, in the direct sum (\ref{eq25}) each of
subspaces ${\cal F}_\lambda$ is invariant under
the enveloping algebra $U[\vec L^{(\lambda)}]$ and,
hence, $su^c(2)$-invariantly (in ${\cal F}_\lambda$ generators of
$su^c(2)$ are the operators $(\vec L^{(\lambda)}\,,i\vec L^{(\lambda)})$,
as the semispinor representation $(\lambda, 0)^+=[\lambda,\lambda +1]^+$ of
the algebra $su^c(2)$, called analytical one, is realized in
${\cal F}_\lambda$ as the representation $\oplus_{p\in{\bf Z}}[\lambda+
\frac{p}{2},\lambda +\frac{p}{2}+1]^+$ in ${\bf F}_{\zeta}$).
A complexification of the algebra $su(2)$ is important in the connection
with a problem of relativization of spin \cite{f3}.
Semispinor representations $(\lambda,0)^+$ of the algebra $su^c(2)$ are
representations spontaneously breaking symmetry of the group $SU^c(2)$
as a whole. Indeed, we shall now see that the Lie algebra
$(\vec L^{(\lambda)}\,,i\vec L^{(\lambda)})$ generates not
representations of the group $SU^c(2)$, but only ones of its open
subgroups connected with the Gauss decomposition $N_+HN_-$ of this group.

      As is known, any regular element $v=\pmatrix{\alpha&\beta\cr
\gamma&\delta}\, (\delta\ne 0)$ of the group $SL(2,{\bf C})$
allows the decomposition $v=n_+hn_-$ where
\begin{displaymath}
n_+=\pmatrix{1&\beta/\delta\cr 0&1}=\exp(\frac{\beta}{\delta}
\sigma_+),\quad h=\pmatrix{\delta^{-1}&0\cr 0&\delta}=
\exp(-\sigma_3\,\ln\delta),
\end{displaymath}
\begin{displaymath}
n_-=\pmatrix{1&0\cr \gamma/\delta&1}=
\exp(\frac{\gamma}{\delta}\sigma_-),
\end{displaymath}
and $\sigma_+=\pmatrix{0&1\cr 0&0},\,\sigma_-=\pmatrix{0&0\cr 1&0},\,
\sigma_3=\pmatrix{1&0\cr 0&-1}$. If to remove singular elements
$\pmatrix{\alpha&\beta\cr -\beta^{-1}&0}\in\Delta'$ from
$SL(2,{\bf C})$, $SL(2,{\bf C})$ will break up on two open
Borel subgroups $B_+=N_+H\ni\pmatrix{\delta^{-1}&\beta\cr 0&\delta}$
and $B_-=HN_-\ni\pmatrix{\delta^{-1}&0\cr \gamma&\delta}$ which are
crossed through the subgroup $H$ of diagonal matrices forming
the Gaussian area $B_+B_-$. The matrices
$\sigma_+\,,\sigma_-\,,\frac{1}{2}\sigma_3$ are generators of
subgroups $N_+\,,N_-\,,H$.
The representations in question are given by mapping
(homomorphism) $T_\lambda:\,\frac{1}{2}\vec\sigma\to\,^{(p)}\!
\bbar\varphi\frac{1}{2}\vec\sigma\varphi=\vec L^{(\lambda +
\frac{p}{2})}$ where $\varphi,\,^{(p)}\!\bbar\varphi$
are defined by the formulae (\ref{eq25}), (\ref{eq26}), and
$\vec L^{(\lambda +\frac{p}{2})}$ are by the formulae (\ref{eq31}).
The operators $\vec L^{(\lambda)}$ set the semispinor
representation $D^+(\lambda)$ of the algebra $su(2)$ (or the representation
$(\lambda,0)^+$ of the algebra $su^c(2)$) in the space ${\cal F}_\lambda$.

      Consider first the subgroup $B_+$. As $T_\lambda$ is a
homomorphism, we have $T_\lambda (b_+)=T_\lambda (n_+)T_\lambda (h)$
where \cite{2e}
\begin{equation}
T_\lambda (h) =e^{-T_\lambda (\sigma_3)\ln\delta}=
e^{-2L_3^{(\lambda)} \ln\delta},\quad
T_\lambda (n_+)=e^{T_\lambda (\sigma_+)\frac{\beta}{\delta}}=
e^{\frac{\beta}{\delta} L_+^{(\lambda)}}=
e^{\frac{\beta}{\delta}\zeta}\,.
\label{eq37}
\end {equation}
As $L_3^{(\lambda)}\zeta^n=(n-\lambda)\zeta^n$ (see (\ref{eq31})),
\begin{equation}
T_\lambda (b_+)f(\zeta)=\delta^{2\lambda}
e^{\frac{\beta}{\delta}\zeta}\,f(\frac{\zeta}{\delta^2})
\label{eq38}
\end{equation}
where $f(\zeta)\in {\cal F}_\lambda$. From (\ref{eq38}) it follows
that a class of holomorphic functions of a complex variable $\zeta$
of order $\rho\leq 1$ and type $0\leq\tau < \infty$ is invariant
under the action of operators $T_\lambda (b_+)$. This class
allows a topologization transforming it into the space of generalized
functions of exponential type $\Phi'$ (see a definition of the
space $\Phi'$ in \cite{2j}). By definition, we have
$\Phi'=\overline{U[L_+^{(\lambda)}]f_0^{(\lambda)}}^{\,\tau_{\Phi'}}$
where $f_0^{(\lambda)}=1$ is the lowest Cartan-Weyl vector of
the representation $D^+(\lambda)$ (a cyclic vector for
$U[L_+^{(\lambda)}]$), and $\tau_{\Phi'}$ is a topology of the space
$\Phi'$ determined in \cite{2j}.

      The space $\Phi'$ is determined so that a $B_+$-orbit of any
vector $f\in\Phi'$ (including any vector of the Cartan-Weyl basis
$f_n^{(\lambda)}$) wholly lies in $\Phi'$, as
$e^{\tau L_+^{(\lambda)}}f(\zeta)=e^{\tau\zeta}f(\zeta)\in\Phi'\,
(\vert\tau\vert <\infty)$. In $\Phi'$ a representation of the (solvable)
group $B_+$ is incompletely reducible, as in $\Phi'$ there exists
a compositional series $\Phi'=\Phi_0'\supset\Phi_1'\supset\cdots
\supset\Phi_N'\supset\cdots\supset 0$ where $\Phi_N'=\overline
{\oplus^\infty_{n=N}f^{(\lambda)}_n}^{\,\tau_{\Phi'}}$ are
$B_+$-invariant subspaces. The operators $T_\lambda (b_+)$
set (in ${\cal F}_\lambda$) the exact representation of the universal
covering $\stackrel{-}{B}_+=N_+\!\!\stackrel{-}{H}$ where
$\stackrel{-}{H}$ is an universal covering for $H$.

      We now turn to subgroup $B_-$. As
\begin{equation}
e^{\tau L_-^{(\lambda)}}\zeta^n=n!\left(-\frac{\tau}{2}
\right)^n L_n^{(-2\lambda-1)}\left(\frac{\zeta}{\tau}
\right)
\label{eq39}
\end{equation}
($L_n^{(\lambda)}$ is Laguerre polinomials \cite{6}), in the linear
system $l.s.\lbrace f_n^{(\lambda)}\rbrace$ there is a flag from
finite dimensional subspaces: $0\subset\Phi^0\subset\Phi^1
\subset\cdots\subset\Phi^N\subset\cdots\subset U[L^{(\lambda)}_+]
\Phi^0$ where $\Phi^N=\oplus^N_{n=0}f^{(\lambda)}_n$ are
$B_-$-invariant subspaces. Hence, an incompletely reducible
representation $B_-\to T_\lambda (b_-)$ of the group $B_-$ is realized
in $l.s.\lbrace f_n^{(\lambda)}\rbrace$.
It is not difficult to obtain that $l.s.\lbrace f_n^{(\lambda)}\rbrace$
allows a $B_-$-invariant closure up to the space $\Phi$ of
trial functions of exponential type containing holomorphic
functions of a complex variable $\zeta$ of order $\rho <1$ and type
$0\leq\tau <\infty$ (see a definition of spaces $\Phi$ in
\cite{2j}). Indeed, from the definition of $\Phi'$ as
a space of linear continuous functionals on the space $\Phi$
\begin{equation}
f(\varphi)=\langle f,\varphi\rangle_\lambda
\label{eq40}
\end{equation}
where $\langle\cdot,\cdot\rangle_\lambda$ is defined by the formula
(\ref{eq33}), and from properties of $\vec L^{(\lambda)}$-invariance of
this functional it follows that $\langle\Phi',e^{\tau
L_-^{(\lambda)}}\Phi\rangle=\langle e^{\bar\tau L_+^{(\bar\lambda)}}
\Phi',\Phi\rangle$. As for any finite $\tau$ we have
$e^{\bar\tau L_+^{(\bar\lambda)}}\Phi'\subset\Phi'$,
$e^{\tau L_-^{(\lambda)}}\Phi\subset\Phi$.

      It is interesting to observe that so-called coherent states
$f^{(\lambda)}_\sigma (\zeta)=\,_0\!F_1(-2\lambda;-\tau\zeta)$
satisfying the equation
$L_-^{(\lambda)}f^{(\lambda)}_\sigma (\zeta)=\sigma
f^{(\lambda)}_\sigma (\zeta),\,\sigma\in{\bf C}$ belong to the
space $\Phi$. On $\Phi$ the operators $T_\lambda (b_-)$ set
the exact representation of the universal covering
$\stackrel{-}{B}_-=\stackrel{-}{H}\!\! N_-$.

      Further, it is clear that the operators $T_\lambda (N_+)$ are
not determined on the space $\Phi$~. As any vector $f\in\Phi$ is
removed at once by the operator
$T_\lambda (n_+)=e^{\zeta\frac{\beta}{\delta}}\,(\beta\neq 0)$
from $\Phi$. In particular it takes place the formula
\begin{displaymath}
T_\lambda (v)\zeta^n=n!\delta^{2\lambda}
\left(-\frac {\gamma}{2\delta}\right)^nL_n^{(-2\lambda-1)}
\left(\frac{\zeta}{\gamma\delta}\right)
e^{\frac{\beta}{\delta}\zeta}
\end{displaymath}
following from (\ref{eq39}) and (\ref{eq38}). On the other hand,
any vector from $\Phi'$ not belonging to the subset
$\Phi\subset\Phi'$ can be removed by the operator $T_\lambda (n_-)$
from $\Phi'$. Indeed, from the formula
\begin{displaymath}
T_\lambda (n_-)e^{\zeta\tau}=(\gamma\tau +1)^{2\lambda}
e^{\zeta\frac {\tau}{\tau\gamma +1}}
\end{displaymath}
or from the more general formula $T_\lambda (n_-)\Phi e^{\zeta\tau}=\Phi
T_\lambda (n_-)e^{\zeta\tau}$ it follows that for $\gamma=-\frac{1}
{\tau}$ the vector $T_\lambda (n_-)e^{\zeta\tau}\not\in\Phi'$.
Thus, on $\Phi'$ there is no representation (as linear one) of the
group $N_-$ in the usual sense.

      Actually, a non-von Neuman representation (see its definition
in \cite{2k}) of the group of one-dimensional chains $\stackrel{\sim}{N}_-$
(and also the group of one-dimensional chains
$\stackrel{\sim}{SL}(2,{\bf C})$, see its definition in \cite{2l})
which certainly is not any manifold (group $N_\mp$ as well as
$SL(2,{\bf C})$ is simply connected) is realized in $\Phi'$.
Thus, remaining in the framework of
the topological vector space $\Phi'$~, it will be impossible
to close the area $B_+B_-$ to obtain the group $SL(2,{\bf C})$
and its representation in $\Phi'$. From the formula
\begin{equation}
T_\lambda(v) e^{\zeta\tau}=(\gamma\tau+\delta)^{2\lambda}
e^{\zeta\frac{\alpha\tau+\beta}{\gamma\tau+\delta}}\,,\quad
v=\pmatrix{\alpha&\beta\cr \gamma&\delta}\in SL(2,{\bf C})
\label{eq41}
\end{equation}
it follows that in $\Phi'$ we have representations of the group of
one-dimensional chains $\stackrel{\sim}{SL}(2,{\bf C})$ not satisfying
the von Neuman axioma about existence of a general (dense
in $\Phi'$) domain of definition for all the operators $T_\lambda (v)$.
From (\ref{eq41}) it follows that the area is trivial, i.\ e.\
$\cap_{v\in SL(2,{\bf C})}D_{T_\lambda (v)}=0$; therefore, corresponding
representations were called by non-von Neuman ones \cite{2k}. Non-von Neuman
representations of topological groups are a subject of the special
discussion which we shall not touch upon here.

      So, the non-Fock representation of the algebra $h_4$ builded in
the dual pair of topological vector spaces $(\Phi'\,,\Phi)$ is generated
such a representation of its group of automorphisms $Sp(2,{\bf R})$
which breaks this group symmetry: a nature of the representation
is such that the complete symmetry is spontaneously lowered up to the
symmetry of open subgroups $B_+$ and $B_-$ connected with the Gauss
decomposition, and the group $B_+=N_+H$ can be represented only in the
space $\Phi'$, and $B_-=HN_-$ is only in $\Phi$. The operators
$L_\pm=L_1\pm iL_2$ and $\vec\Gamma\pm i\vec N$ are generators
of subgroups $N_\pm$,
and $L_3$ and $\Gamma_0$ are generators of $H$. Thus, semispinors
are objects with a restricted symmetry. And there are two kinds of
semispinors: semispinors having symmetry of the group $B_+$
(associated with the Gauss decomposition $N_+HN_-$ and the pair of spaces
$(\Phi'\,,\Phi)$) and semispinors having symmetry of group $B_-$
(assiciated with another Gauss decomposition $N_-HN_+$ and another pair of
spaces $(\Phi,\Phi')$). Both both the pairs of spaces and both the
Gauss decompositions are connected by the time reflection operator $T$
for which formulae are read from the right on the left. So only at the level
of semispinors we have an opportunity to break both $T$-symmetry
and $CP$-symmetry which are impossible to break
at a level of spinors and spinor fields (see \cite{7}).

\section*{V. Extended Fock representation of the algebra $h_4$.
Additional variables}\label{kd5}
      1) The non-Fock representation $T_1(h_4)$ proves to be
in some sense by an expansion of the Fock representation $T_0(h_4)$
for the same algebra, i.\ e.\ $T_0(h_4)\subset T_1(h_4)$.

\begin{aff}\label{San4}
The representation $T_1(h_4)$ is interlaced with the extended Fock
representation of the algebra $h_4$ which, as well as the
Fock representation,
is given by the operators $a^1_\alpha={\partial}/{\partial z_\alpha}\,,
a^2_\alpha=z_\alpha\,(\alpha=1,2)$, but is realized in the space
${\bf F}={\cal F}_F\otimes {\cal F}_0$ where ${\cal F}_F$ is a
space of the Fock representation $T_0(h_4)$ formed
by functions of two complex variables $z_1\,,z_2$, and
${\cal F}_0$ is a space of functions depending on an additional
variable $z\dot =z_2$ (in this case ${\partial}/{\partial z}
\dot={\partial}/{\partial z_2}$) having non-standard
transformational properties.
\end{aff}

      The existence of additional variables in $T_1(h_4)$
is connected to the existence of two various units on the space
${\bf F}_\zeta^{(\lambda)}$. This becomes obviously if
${\bf F}_\zeta^{(\lambda)}$ to write down as a column
\begin{equation}
{\bf F}_\zeta^{(\lambda)}=\left\lgroup\matrix{\vdots\cr
{\cal F}_{\lambda +\frac{1}{2}}\cr{\cal F}_\lambda\cr
{\cal F}_{\lambda-\frac{1}{2}}\cr\vdots\cr}\right\rgroup
\label{eq42}
\end{equation}

      Then one of units which we shall denote by
$\mbox{\Large\bf 1}$ is represented
by unity diagonal matrix infinite in all parties
\begin{equation}
\mbox{\Large\bf 1}=\left\lgroup\matrix{\ddots&&&&\cr
&1&&\ \mbox{\huge 0}&\cr &&1&&\cr &\!\!_{\mbox{\Huge 0}}&&1&\cr
&&&&\ddots\cr}\right\rgroup
\label{eq43}
\end{equation}
(a warning: this unit should not be mixed up with the unit of
the basic field of scalars, which enters in the right part of
commutation relations (\ref{eq22}) and under which it follows
to understand $\mbox{\Large\bf 1}^{\!0}$). In (\ref{eq43}) the unit 1 as
an operator on the subspace ${\cal F}_\lambda$ will be invariant under
transformations from the group $SU(2)$ \cite{2e,2f}:
\begin{equation}
T_\lambda (u)\cdot 1\cdot T^{-1}_\lambda (u)=1\,,\quad
u=\pmatrix{\alpha&\beta\cr -\bar\beta&\bar\alpha}\in
SU(2)
\label{eq44}
\end{equation}
where the operators $\vec L^{(\lambda)}$ (\ref{eq31}) are generators
of $T_\lambda(u)$. Thus, the unit
$\mbox{\Large\bf 1}$ is an $SU(2)$-scalar (it is
a scalar under transformations of $U(2)$ and $U^c(2)$ too).

      Another unit is the second component of the
spinor $a^1_k$ (see the formulae (\ref{eq21}), (\ref{eq25})). And
if the space ${\bf F}_\zeta^{(\lambda)}$ is written down as
(\ref{eq42}), this unit is represented by a matrix in which
units are displaced (in comparison with the matrix (\ref{eq43}))
on one step in the right top corner
\begin{equation}
a^1_2=\left\lgroup\matrix{\ddots&\ddots&&&&\cr
&0&1&&\ \mbox{\huge 0}&\cr &&0&1&&\cr &\!\!_{\mbox{\Huge 0}}&&0&1&\cr
&&&&\ddots&\ddots\cr}\right\rgroup
\label{eq45}
\end{equation}
The matrix too is possible to consider as diagonal one: you see,
in the case the concept of diagonal is rather conditional (the
first element is not in (\ref{eq42})). So both matrices
$\mbox{\Large\bf 1}$ and $a^1_2$ are actually indistinguishable.
However in the case of $a^1_2$ the unit 1 in (\ref {eq45}) as a
operator on ${\cal F}_\lambda$ will be transformed absolutly under
another law, namely, \cite{2e,2f}:
\begin{equation}
T_{\lambda-\frac{1}{2}}(u)\cdot 1\cdot T^{-1}_\lambda(u) =
-\bar\beta\frac{d}{d\zeta}+\bar\alpha\cdot 1
\label{eq46}
\end{equation}
following from (\ref{eq26}). An equality of values having
different transformational properties denotes as
\begin{equation}
a^1_2\dot =\mbox{\Large\bf 1}\quad .
\label{eq47}
\end{equation}
An analogous situation takes place in the case of Dirac matrices
$\beta\dot =\gamma_4$ where $\beta$ enters into the definition of the
Dirac conjugate bispinor $\bar\psi=\psi^+\beta$ and
is transformed under the formula $S^+\beta S=\beta$, while
$\gamma_4$ being by the fourth component of a vector $\gamma_\mu$
will be transformed under the law $S^{-1}\gamma_4 S=L_{4\mu}\gamma_\mu$.
The operators
\begin{displaymath}
\left\lgroup\matrix{\ddots&&&&\cr &0&&\ \mbox{\huge 0}&
\cr &&1&&\cr &\!\!_{\mbox{\Huge 0}}&&0&\cr &&&&\ddots\cr}\right\rgroup
\end{displaymath}
are marked, too, by the ambiguity specified above. This means that in the
non-unitary theory there are neither projectors nor any quantum logic
(compare with \cite{3}).

      If now to pass from the $\zeta$-realization (\ref{eq25}) to
the Fock $z$-realization with the help of the transformation \cite{2e}
\begin{equation}
f(\zeta)=\int K(\zeta;\bbar z_1\,,\bbar z_2)\,f(z_1\,,z_2)\,
d\mu (z_1\,,z_2)=\stackrel{\wedge}{K}f(z_1\,, z_2)
\label{eq48}
\end{equation}
where $d\mu(z_1\,,z_2)$ is the Gauss measure on ${\bf C}^2$
(see (\ref{eq8})), and (compare the structure $K$ with the structure of
the space ${\bf F}_\zeta$ (\ref{eq25}))
\begin {equation}
K(\zeta;\bbar z_1\,,\bbar z_2)=\bigoplus_{p\in {\bf Z}}
\bbar z_2^{2\lambda +p}e^{\zeta\frac {\bbar z_1}{\bbar z_2}}
\label{eq49}
\end{equation}
(The operator $\stackrel{\wedge}{K}$ in (\ref{eq48}) possesses
the property
\begin{equation}
\stackrel{\wedge}{K}a_\alpha^{\rm a}(z)=a_\alpha^{\rm a}(\zeta)
\stackrel{\wedge}{K}
\label{eq50}
\end{equation}
and refers to as interlacing one. Here $a_\alpha^{\rm a}(z)$ are defined
by the formula (\ref{eq27}), and $a_\alpha^{\rm a}(\zeta)$ are by the
formula (\ref{eq25})), by virtue of the diagram (here in
accordance with (\ref{eq48}) we have written down $\stackrel
{\wedge}{K}$ as $\stackrel{\wedge}{K}=K\circ d\mu$
and have taken into account that $\bbar z_\alpha\,d\mu=-\frac{\partial}
{\partial z_\alpha}\,d\mu$\,, and $T$ is the mapping of a cotangent
space in a tangent space)
\begin{equation}
\def\mapright#1{\smash{\mathop{\rightarrow}\limits^{#1}}}
\def\mapleft{\smash{\mathop{\leftarrow}\limits^T}}
\def\mapup#1{\Big\uparrow\rlap{$\vcenter{\hbox{$\scriptstyle#1$}}$}}
\matrix{a^1_1(\zeta)&\mapright T&a^2_1(\zeta)&&&&a^2_2(\zeta)&
\mapleft&a^1_2(\zeta)&\dot =&\mbox{\Large\bf 1}&\mapright {}&?\cr
\mapup K&&\mapup K&&&&\mapup K&&\mapup K&&&&\mapup K\cr
\bbar z_1&\mapright T&\partial/\partial\bbar z_1&&&&\partial/\partial\bbar z_2&
\mapleft&\bbar z_2&\dot =&\bbar z&\mapright {}&\partial/\partial\bbar z\cr
\mapup {d\mu}&&\mapup {d\mu}&&&&\mapup {d\mu}&&\mapup
{d\mu}&&\mapup {d\mu}&&\mapup {d\mu}\cr
\partial/\partial z_1&\mapleft&z_1&&&&z_2&\mapright T&\partial/\partial z_2&
\dot =&\partial/\partial z&\mapleft&z\cr}
\label{eq51}
\end{equation}
the equality (\ref{eq47}) will cause existence of the variable
$z\dot =z_2$ being by an $SU(2)$-scalar, as well as $\mbox{\Large\bf 1}$.
\begin{aff}
If $f(\bbar z_2$ is a function, $f(\bbar z_2)\,K=C\,K$
where $K$ is defined by the formula (\ref{eq49}), and $C=f(1)$\cite{f4}.

      It follows from this that $K$ has a non-trivial kernel:
${\rm Ker}\,K=\lbrace f(\bbar z_2)\vert f(1)=0\rbrace$, therefore
$\stackrel{\wedge}{K}$ does not have the inverse (so,
indeed, $\stackrel{\wedge}{K}$ is a interlacing operator).
\end{aff}
\begin{aff}
The factor space ${\bf F}_\zeta^{(\lambda)}/{\rm Ker}\,K$
is isomorphic to the space ${\bf F}_\zeta^{(\lambda)}$. Hence,
we have ${\cal F}_F\subset{\bf F}_\zeta^{(\lambda)}\subset{\bf F}_z$.
\end{aff}
{\bf Consequence }{\it The non-standard oscillator is equivalent
to the double standard oscillator described by variables
$z_2\,,{\partial}/{\partial z_2}\quad{\it and}\quad z\dot =z_2\,,
{\partial}/{\partial z}\dot ={\partial}/{\partial z_2}$.}

      This statement is almost a trivial consequence of the Zermelo theorem
(or the axioma of choice): it follows from an opportunity of complete
ordering of the set ${\bf Z}=\lbrace\ldots,-2,-1,0,1,2,\ldots
\rbrace$, which is written down as
${\bf Z}=\lbrace 0,1,2,\ldots;-1,-2,\ldots\rbrace=\lbrace{\bf Z}_+\,,
\overline{\bf Z}_+\rbrace$ where ${\bf Z}_+=\lbrace 0,1,2,\ldots\rbrace$
and $\overline{\bf Z}_+=-\lbrace{\bf Z}_+\!\!\!\setminus\! 0\rbrace$.
Here as states of the standard oscillator number by the set ${\bf Z}_+$
as states of the second oscillator by the set $\overline{\bf Z}_+$
(states with negative occupation numbers) which are described states
of physical vacuum. However the fact that the second oscillator has
trivial transformational properties is a non-trivial result at all.

      It is necessary to notice that the physical
vacuum (as a ground state of the World) was postulated
by Heisenberg in his time \cite{9}.

      Thus, it is possible to write down that
$T_1(h_4)\sim T_0(h_4)+T_0(h_2)$, and the space of the extended
Fock representation to write down as
${\bf F}=\overline{U[z_1\,, z_2]{\cal F}_0}^{\,\tau}$ where the space
of additional variables ${\cal F}_0$ plays a role of a cyclic
subspace (as is known, the space of the Fock representation is generated
by a unique cyclic vector, by the mathematical vacuum 1, therefore
${\cal F}_F=\overline{U[z_1\,,z_2]\cdot 1}^{\,\tau}$).

      In general we shall have $T_k(h_{2n})\sim T_0(h_{2n})+T_0(h_{2k})$
where $h_{2k}$ is a subalgebra in $h_{2n}$,
so $T_k(h_{2n})$ contains $k$ additional variables.

      It is interesting to notice that the existence of an additional
variable, the $SU^c(2)$-scalar $z\dot =z_2$, was felt for a long time.
It is enough to look at the formulae which set representations of the
group $SU(2)$ in the class of functions depending on the projective variable
$z_1/z_2$ (for example, see \cite{10}). The pure geometrical proof
of existence of an $SU(2)$-scalar is given in \cite{2k}. However here we are
not stopped on the proof.

      2) In the propounded theory the additional variables play a
role of hidden parameters which for a long time and unsuccessfully try
to enter into the quantum theory (an introduction of them is forbidden
by the well-known von Neuman theorem \cite{3}). Nevertheless an
introduction of additional variables proves to be possible. However,
it is connected with a radical reorganization of the
Heisenberg-Schr$\rm\ddot o$dinger quantum theory: it turns from the
unitary theory in non-unitary one. Moreover a sense of
hidden variables is absolutly other: they are not additional variables
of usual quantum objects (as they were understood by de Broglie \cite{11}),
and variables of the absolutly other physical object --- the dynamical
system generating usual quantum objects (elementary
particles). And it will be shown further.

      If to write down the formula of spaces ${\bf F}_\zeta^{(\lambda)}=
\stackrel{\wedge}{K}\!\!\!({\cal F}_F\otimes{\cal F}_0)$
interlaced by the operator $\stackrel{\wedge}{K}$  for
representations of the group $SU(2)$, which is realized in these spaces,
we shall obtain the relation (compare with (\ref{eq50}))
\begin{equation}
\left(\bigoplus_{p\in {\bf Z}}D^+(\lambda +\frac{_p}{^2})\right)
\!\!\stackrel{\wedge}{K}=\stackrel{\wedge}{K}\!\!
\left(\bigoplus_{n\in{\bf Z}_+}D(\frac{_n}{^2})\otimes D_h(0)\right)
\label{eq52}
\end{equation}
where $D(\frac{n}{2})$ are finite dimensional representations realized
in the Fock space ${\cal F}_F$, and $D_h(0)$ is the non-trivial
unity representation realized in the space of additional
variables ${\cal F}_0$. It is necessary to pay attention on the fact that
the relation (\ref{eq52}) is carried out at any $\lambda$ with the
exception of integer or half-integer numbers.

       As is proved in \cite{2m}, the semispinor representation
$D^+(\lambda)\ (\lambda\ne\frac{n}{2})$ of the algebra $su(2)$
is irreducible algebraically and topologically . Hence, it is not
equivalent to the well-known finite dimensional representations of
this algebra. However the infinite system of semispinor representations
is equivalent to an infinite system of spinor representations (the
formula (\ref{eq52})). This fact is of fundamental importance for
physics: the formula (\ref{eq52}) describes how
the complete symmetry of $SU(2)$ and $SU^c(2)$ (inherent to
finite dimensional objects --- spinors) arises of the spontaneously
broken symmetry inherent to semispinors --- infinite dimensional
objects (see section~IV).

      However, in spite of the symmetry of the group $SU(2)$ being
restored, the complete symmetry of the group $Sp(2,{\bf R})$
in the space of the extended Fock representation remains still
broken. The breaking is caused by a topologization of representation space
( to put it more exactly, by the topology of the dual pair of spaces
$(\Phi'\,,\Phi)$): a representation of the subgroup $e^{i\vec\theta
(\vec\Gamma +i\vec N)}$ is realized only in the space of generalized
functions of exponential type $\Phi'$, while a representation of
another subgroup $e^{i\vec\theta (\vec\Gamma-i\vec N)}$ is in $\Phi$.
It is interesting to notice that if to reject the non-trivial unity
representation $D_h(0)$ we shall come to the unitary theory. From this
it follows such a conclusion that from the point of view of the non-unitary
theory the unitary theory is not complete. The latter gives the description
of quantum objects only, while the non-unitary theory gives the joint
description of quantum and subquantum objects (the subquantum object
is understood to be the physical vacuum related to the space ${\cal F}_0$).

      3) It is interesting also to notice that the additional variable
is associated with splitting of the phase transformation
$U(1)=e^{i\alpha L_0}$. Indeed, on the one hand, it should be $L_0 z=0$, as
$z$ is an $SU(2)$-scalar, and, hence, $\vec L z=0$ and $\vec L^2 z=L_0
(L_0 + 1)z=0$. The operator $L_0$ is denoted by $L^{(l)}_0$.
On the other hand, $z$ cannot be an $(\vec N,\vec \Gamma)$-scalar, as
conditions $\vec N z=\vec\Gamma z=0\,\ \Gamma_0 z=\frac{1}{2} z$ are not
joint with values of the Casimir operators $\vec L^2-\vec N^2=-\frac{3}{4}\,\ \vec
\Gamma^2-\Gamma_0^2=\frac{1}{2}$~. $z$ must behave under these
transformations as $z_2$, i.\ e.\ (as $z\dot=z_2$) it should be
$L_0 z=\frac{1}{2}z$. The operator $L_0$ is denoted by
$L^{(i)}_0$. Thus, we have $L^{(l)}_0 z=0$ and $L_0^{(i)}z=\frac{1}{2}z$.
In this connection it should be to put that
$\vec L^2=L_0^{(l)}(L_0^{(l)}+1),\ \vec N^2=L_0^{(l)}(L_0^{(l)}+1)+
\frac{3}{4}$ and $\Gamma_0=L_0^{(i)}+\frac{1}{2}\,\ \vec\Gamma^2=
(L_0^{(i)}+\frac{1}{2})^2+\frac{1}{2}=L_0^{(i)}(L_0^{(i)}+1)+\frac{3}{4}$.
The contradiction specified above is completely removed.

      It is resonable to attach the phase transformation
$e^{i\alpha L_0^{(l)}}$ to $e^{i\vec L\vec\theta}$ by expanding, thus,
the group $SU(2)$ up to $U(2)$ and to add the transformation
$e^{i\vec\alpha\vec\Gamma}$ by the transformation $e^{-i\alpha_0\Gamma_0}$
up to the transformation $e^{i\alpha_\mu\Gamma_\mu}$. Note that an analogous
splitting of phase transformation was postulated by Heisenberg \cite{9}
in his theory of a ground state of the World.

\section*{VI. Non-Fock representation $T_2(h^{(*)}_8)$ of
the algebra $h^{(*)}_8$}
      1) The algebra $h_4$, appeared for the first time in Majorana's work
\cite{5} in connection with the known equations that now bear his name
(see also \cite{12}), is unsufficient from the physical point of view
for many reasons. First, though the theory of representations of this
algebra allows to build relativistic (but Majorana's) 4-spinors
under the formula $\psi_k=\langle\dot f,\phi_k f\rangle$ where
$\phi_k={a^1_k\choose a^2_k}$, however it does not allow to obtain
fields $\psi (X)$ on the space-time ${\bf A_{3,1}}$ if not
to postulate their existence of the space ${\bf A_{3,1}}$.
Certainly, it is a serious lack. And the point is that the group of
automorphisms $Sp(2,{\bf R})$ of the algebra $h_4$, locally isomorphic
to the de Sitter group $SO(3,2)$ which acts in the Majorana fiber
$S_4\ni\psi$, contains neither the Poincar$\rm\acute e$ group
${\cal P}$ nor the translation group with which only it
would be possible to connect (not to postulate) the
existence of the space ${\bf A_{3,1}}$ (nevertheless Dirac
particles are necessary rather than Majorana ones in particle
theory). Secondly, a representation of the subalgebra $sl(2,{\bf C})$
(generated in ${\bf F}_\zeta^{(\lambda)}$ by the operators $\vec L\,,\vec N$,
see section~III), as well as a representation of the algebra
$sp(1,{\bf C})$ in the space ${\cal F}_F$ (see section~II)
is a cycle of two representations $[\frac{1}{2}\,,0]^+\oplus
[0,\frac{1}{2}]^+$ which do not form a complete system.
The decycling results in an infinite complete system of
real semispinor representations of the group $SU^c(2)\approx
SL(2,{\bf C})$ of the following (an $SU^c$-analogue of the formula
(\ref{eq30}) \cite{f5})
\begin{equation}
\bigoplus_{p\in {\bf Z}}\bigoplus_{q\in{\bf Z}}
(\lambda+\frac{_p}{^2}\,,\lambda'+\frac{_q}{^2})^+
\label{eq53}
\end{equation}
(real representations $(\lambda\,,\lambda')=(\lambda\,,0)^+\,
+\,\overline{(\bar\lambda'\,,0)^+}$ where $\overline{(\bar\lambda'
\,,0)^+}=(0,\lambda')$ is an antianalytical spinor representation
are investigated in \cite{2n,2o}). The space of representation (\ref{eq53})
is written down analogously (an analogue of the formula (\ref{eq25}))
\begin{equation}
{\bf F}^{(\lambda\,,\lambda')}_{\zeta\,,\bar\zeta}=
\bigoplus_{p\in{\bf Z}}\bigoplus_{q\in {\bf Z}}
{\cal F}_{\lambda +\frac{p}{2}\,,\lambda'+\frac{q}{2}}\quad.
\label {eq54}
\end{equation}
Now together with the operators $\varphi^\alpha\,,\,^{(p)}\!\bbar
\varphi_\alpha$ (\ref{eq25}) having properties
\begin{displaymath}
\varphi^\alpha:\,{\cal F}_{(\lambda +\frac{p}{2}\,,\lambda'+\frac{q}{2})}
\to {\cal F}_{(\lambda +\frac{p-1}{2}\,,\lambda'+\frac {q}{2})}\,;\quad
^{(p)}\!\bbar\varphi_\alpha:\,{\cal F}_{(\lambda +\frac{p}{2}\,
,\lambda'+\frac{q}{2})}\to{\cal F}_{(\lambda +\frac{p+1}{2}\,
,\lambda'+\frac{q}{2})}
\end{displaymath}
it is necessary to consider also the complex conjugate operators
$\varphi^{*\,\dot\alpha}\,,\,^{(q)}\!\bbar\varphi^*_{\dot\alpha}$
(spinors with dotted indices) having properties
\begin{displaymath}
\varphi^{*\,\dot\alpha}:\,{\cal F}_{(\lambda +\frac{p}{2}\,
,\lambda'+\frac{q}{2})}\to{\cal F}_{(\lambda +\frac {p} {2} \,
,\lambda'+\frac{q-1}{2})}\,;\quad
^{(q)}\!\bbar\varphi^*_{\dot\alpha}:\,{\cal F}_{(\lambda +\frac{p}{2}\,
,\lambda'+\frac{q}{2})}\to{\cal F}_{(\lambda +\frac{p}{2}\,
,\lambda'+\frac{q+1}{2})}
\end{displaymath}
where $*$ is the complex conjugation. By analogue with (\ref{eq21})
we shall enter in consideration of the operators $a^{\rm a}_\alpha$
and $a^{\rm a\,*}_\alpha$. Their restrictions on subspaces
${\cal F}_{(\lambda +\frac{p}{2}\,,
\lambda'+\frac{q}{2})}={\cal F}_{\lambda +\frac{p}{2}}\times
\overline{{\cal F}_{\bar\lambda'+\frac{q}{2}}}$ are
the  operators $\varphi\,,\bbar\varphi\,,\varphi^*\,,
\bbar\varphi^*$ entered just now, and we shall consider the operators
\begin{equation}
\phi={a_\alpha^{2\,*}\choose a^1_\alpha},\quad
\bar\phi=\left(-a_\alpha^{1\,*},\,a^2_\alpha\right).
\label{eq55}
\end{equation}
They satisfy commutation relations (they follow
from commutation relations for $a$ and $a^*$)
\begin{equation}
[\phi_\alpha\,,\bar\phi_\beta] =\delta_{\alpha\beta}\,,
\quad [\phi_\alpha\,,\phi_\beta]=[\bar\phi_\alpha\,,\bar\phi_\beta]=0
\label{eq56}
\end{equation}
which is defined the Heisenberg algebra with involution
$h_8^{(*)}=h_4+h^*_4$~. Its representation in the space
${\bf F}^{(\lambda\,,\lambda')}$ (\ref{eq54}) is denoted by
$T_2(h^{(*)}_8)$ and refers to as non-Fock one. A representation
(connected to it) of the algebra $sl(2,{\bf C})$ (\ref{eq53}) is
quite obviously given
by the operators $I_{\mu\nu}=\bar\phi\Sigma_{\mu\nu}\phi$ where
$\Sigma_{\mu\nu}=\frac{1}{4i}[\gamma_\mu\,,\gamma_\nu]$, and
$\gamma_\mu$ are Dirac matrices. Also it is obvious that the
representation of $sl(2,{\bf C})$ is extended up to a
representation of $gl(2,{\bf C})$ with the help of
the operators $A=\bar\phi\phi$ and $B=\bar\phi\gamma_5\phi$.
Proceeding from commutation relations (\ref{eq56}), we come to
relations
\begin{displaymath}
[I_{\mu\nu}\,,I_{\rho\sigma}]=\delta_{\nu\rho}I_{\mu\sigma}+
\delta_{\mu\sigma}I_{\nu\rho}-\delta_{\mu\rho}I_{\nu\sigma}-
\delta_{\nu\sigma}I_{\mu\rho}\,,
\end{displaymath}
\begin{displaymath}
[I_{\mu\nu}\,,A]=[I_{\mu\nu}\,,B]=[A\,,B]=0\,.
\end{displaymath}

      So the relativization of spin (formula (\ref{eq53})) results in
consideration of the Heisenberg algebra with involution $h_8^{(*)}$
which we was already obtained earlier in \cite{1}. There
the variables $\bar\phi$ were Dirac conjugate to $\phi$~,
i.\ e.\ they were connected with $\phi$ by the formula
$\bar\phi=\phi^+\beta$
where $\beta\dot =\gamma_4=\pmatrix{0&1\cr 1&0}$ is the Dirac matrix,
and $+$ is a conjugation with respect to a sesquilinear form which
it will be necessary now to set on ${\bf F}^{(\lambda\,,\lambda')}$.
For the operators $a^a_\alpha$ a condition of the Dirac conjugation
means that $(a^1_\alpha)^+=-a^{1\,*}_\alpha$, and $a^{2\,+}_\alpha=
a^{2\,*}_\alpha$ or (compare with the formula (\ref{eq9}))
\begin{equation}
\langle\dot f,a^1_\alpha f\rangle=-\langle a^{1\,*}_\alpha
\dot f,f\rangle,\quad
\langle\dot f, a^2_\alpha f\rangle=\langle a^{2\,*}_\alpha
\dot f,f\rangle
\label{eq56'}
\end{equation}
where $\langle\cdot,\cdot\rangle$ is the mentioned above form on
$f\in {\bf F}^{(\lambda\,,\lambda')}$, and
$\dot f\in\dot{\bf F}^{(\lambda\,,\lambda')}$ is the space dual
to ${\bf F}^{(\lambda\,,\lambda')}$ with respect to the form
$\langle\cdot,\cdot\rangle$. The form was determined
in \cite{2o}. It is written down as
\begin{equation}
\langle\dot f,g\rangle=\sum^\infty_{p,q=-\infty}
\langle f^{(-\bar\lambda'-1-\frac{q}{2}\,,-\bar\lambda-1-\frac{p}{2})}
\,,g^{(\lambda +\frac{p}{2}\,,\lambda'+\frac{q}{2})}\rangle
\label{eq57}
\end{equation}
where $g^{(\lambda +\frac{p}{2}\,,\lambda'+\frac{q}{2})}$ is
a projection of $g$ in the subspace
${\cal F}_{(\lambda +\frac{p}{2}\,,\lambda'+\frac{q}{2})}$,
$f^{(-\bar\lambda'-1-\frac{q}{2}\,,-\bar\lambda-1-\frac{p}{2})}=
\dot f^{(\lambda +\frac{p}{2}\,,\lambda'+\frac{q}{2})}$ is
a projection of $f$ in the subspace
${\cal F}_{(-\bar\lambda'-1-\frac{q}{2}\,,-\bar\lambda-1-\frac{p}{2})}=
\dot{\cal F}_{(\lambda +\frac{p}{2}\,,\lambda'+\frac{q}{2})}$,
in which is realized the representation
$(-\bar\lambda'-1-\frac{q}{2}\,,-\bar\lambda-1-\frac{p}{2})^+$
dual to $(\lambda +\frac{p}{2}\,,\lambda'+\frac{q}{2})^+$
( as is shown in \cite{2o}, a pair $(-\bar\lambda'-1\,,-\bar\lambda-1)$
is conjugate to another pair of numbers $(\lambda\,,\lambda')$), and
$\langle\cdot,\cdot\rangle$ is given by the integral
\begin{equation}
\langle\dot f^{(\lambda\,,\lambda')}\,,g^{(\lambda\,,\lambda')}\rangle=
\int\overline{\dot f^{(\lambda\,,\lambda')}(\zeta\,,\bar\zeta)}\,
Ig^{(\lambda\,,\lambda')}(\zeta\,,\bar\zeta)\,d\mu (\zeta)
\label{eq58}
\end{equation}
where $d\mu=\frac{i}{4\pi}d\zeta\wedge d\bar\zeta$ is an
$SL(2,{\bf C})$-invariant measure on ${\bf C}$ and $Ig(\zeta,
\bar\zeta)=g(-\zeta,-\bar\zeta)$\cite{2o}.

     The Hermitian property of operators $p_\mu=i\bar\phi
\gamma_\mu P_+\phi$ and $\dot p_\mu=-i\bar\phi\gamma_\mu P_-\phi$
(they expand the algebra $gl(2,{\bf C})$ up to the algebra $u(2,2)$)
is connected with the condition of Dirac conjugation of variables
$\phi$ and $\bar\phi$ in the metric $\langle\cdot,\cdot\rangle$,
i.\ e.\ $\langle p_\mu\dot f,g\rangle=\langle\dot f,p_\mu g\rangle$
(and similarly for $\dot p_\mu$). However to throw exponents
$e^{ixp}$ (or $e^{i\dot x\dot p}$) from left to right (or from the right
on the left) in $\langle\cdot,\cdot\rangle$ is not allowable because
of the symmetry of the group $Sp^{(*)}(4,{\bf C})$ spontaneously
broken (see \cite{1}, and also section~IV).

      It is easy to see that a special (non-Wigner) representation
of the Lie algebra of the Poincar$\rm\acute e$ group ${\cal P}=
SL(2,{\bf C})\times\!)T_{3,1}$ (likewise for the group
$\dot{\cal P}=SL(2,{\bf C})\times\!)\dot T_{3,1}$ represented
in the dual space $\dot{\bf F}^{(\lambda\,,\lambda')}$) is realized
on the space ${\bf F}^{(\lambda\,,\lambda')}$. First, it is necessary
to notice that here all the generators of the group ${\cal P}$
(and they are $I_{\mu\nu}\,,p_\mu$) are vertical vectors
(acting in a fiber, see \cite{1}), in the Wigner representation
$p_\mu=-i\frac{\partial}{\partial X_\mu}$ are horizontal
vectors (they act in the base). Secondly, it is easy to see that
in $\dot{\bf F}^{(\lambda\,,\lambda')}$ there is a flag from
$p_\mu$-invariant subspaces such that a solvable representation of
the Lie algebra of the subgroup $T_{3,1}$ is realized in
$\dot{\bf F}^{(\lambda\,,\lambda')}$ (the Wigner approach
where there is the base ${\bf A}_{3,1}$ make use of
irreducible representations of the group $T_{3,1}$).

      2) The majority of statements for the non-Fock representation theory
of the algebra $h_8^{(*)}$ copies statements of the algebra $h_4$. So,
as well as in the case of the algebra $h_4$, the non-Fock representation
of the algebra $h_8^{(*)}$ is interlaced with its extended Fock
representation such that relations are fulfilled as
\begin{equation}
\phi (\zeta)\!\!\stackrel{\wedge}{K}=\stackrel{\wedge}{K}\!\!\phi (z)\,,\quad
\bar\phi (\zeta)\!\!\stackrel{\wedge}{K}=\stackrel{\wedge}{K}\!\!\bar\phi(z)
\label{eq59}
\end{equation}
in which the interlacing operator $\stackrel{\wedge}{K}$ is defined
by the formula
\begin{displaymath}
f(\zeta\,,\bar\zeta)=\stackrel{\wedge}{K}\!\! f(z\,,\bbar z)=
\int K(\zeta\,,\bar\zeta\,; z\,,\bbar z)\,\overline{f(z\,,\bbar z)}\,
d\mu (z)\,.
\end{displaymath}
Here
\begin{displaymath}
K(\zeta\,,\bar\zeta\,;z\,,\bbar z)=\left(\bigoplus_{p\in{\bf Z}}
\bbar z_2^{2p+\lambda}\,e^{\zeta\frac{\bbar z_1}{\bbar z_2}}\right)\times
\left(\bigoplus_{q\in {\bf Z}}\bbar z_2^{2q+\lambda'}\,
e^{-\bar\zeta\frac{z_1}{z_2}}\right),
\end{displaymath}
and $d\mu (z)=\left (\frac{i}{4\pi}\right)^2\prod_{\alpha=1,2}
dz_\alpha\wedge d\bbar z_\alpha$ is an $SL(2,{\bf C})$-invariant
measure on the Lagrangian plane $L\ni (z_\alpha\,,\bbar z_\alpha)$.
In (\ref{eq59}) the operators $\phi (\zeta)\,,\bar\phi (\zeta)$ are
written down in the $\zeta$-realization (\ref{eq55}), (\ref{eq25}),
and $\phi (z)\,,\bar\phi (z)$ are in the $z$-realization, in which they
have the same form, as well as in the Fock representation, i.\ e.\
\begin{equation}
\phi={{\partial}/{\partial\bbar z_\alpha}\choose z_\alpha},\quad
\bar\phi=\left (\bbar z_\alpha\,,-\frac{\partial}{\partial z_\alpha}\right),
\quad\alpha=1,2\,,
\label{eq60}
\end{equation}
however, a representation space is the space
\begin{equation}
{\bf F}={\cal F}_F\bigotimes{\cal F}_0
\label{eq61}
\end{equation}
where ${\cal F}_F$ is the Fock representation space
formed by functions of complex variables $z_\alpha\,
,\bbar z_\alpha$, and ${\cal F}_0$ is the space of functions
depending on additional variables $z\dot =z_2\,,\bbar z\dot =
\bbar z_2$ which are $GL(2,{\bf C})$-scalars (the same structure has
the dual space $\dot{\bf F}$). An $SL(2,{\bf C})$-invariant
sesquilinear form on ${\bf F}$, connecting ${\bf F}$
and $\dot{\bf F}$ together in a dual pair of spaces, is defined
by the formula \cite{2o}
\begin{equation}
\langle\dot f,g\rangle=\int\overline{\dot f(z)}\,g(z)\,d\mu (z).
\label{eq62}
\end{equation}
And the variables $\phi$ and $\bar\phi$ (\ref{eq60}) satisfy
relations (\ref{eq56'}) in which now $\langle\cdot,\cdot\rangle$
is understood to be the form (\ref{eq62}).

      An analogue of the formula (\ref{eq52}) will be
\begin{equation}
\bigoplus_{p\in{\bf Z}}\bigoplus_{q\in{\bf Z}}\left (\lambda +\frac{_p}{^2}
\,,\lambda'+\frac{_q}{^2}\right)^+
\!\!\stackrel{\wedge}{K}=\stackrel{\wedge}{K}\!\!
\left (\!\left (\bigoplus_{n\in{\bf Z}_+}\bigoplus_{m\in{\bf Z}_+}
(\frac{_n}{^2}\,,\frac{_m}{^2})\right)\times (0,0)_h\right)
\label{eq63}
\end{equation}
where $(\frac{n}{2}\,,\frac{m}{2})$ are finite dimensional (spinor)
representations of the group $SL(2,{\bf C})$ realized in ${\cal F}_F$
, and $(0,0)_h$ is the non-trivial unity representation realized
in ${\cal F}_0$. Here real semispinor representations
$(\lambda +\frac{p}{2}\,,\lambda'+\frac{q}{2})^+$ (as well as
analytical ones, see section~IV) spontaneously break the
symmetry of $SL(2,{\bf C})$, lowering it up to the symmetry of its
open subgroups, while finite dimensional representations are
characterized by the complete symmetry of $SL(2,{\bf C})$. And
nevertheless despite the fact that in the extended Fock representation
the symmetry of $GL(2,{\bf C})$ is complete, $U(2,2)$-symmetry remains
broken by the topology of the dual pair of spaces $(\Phi'\,,\Phi)$
(compare with a case of algebra $h_4$) that it will be seen from
further considerations. In particular, the exponent $e^{ipx}$ is
connected with the space $\Phi'$, and $e^{i\dot p\dot x}$ is with $\Phi$.

      In the $z$-realization for the operators $p_\mu$ and
$\dot p_\mu$ we have expressions
\begin{equation}
p_\mu=\bbar z\stackrel{+}{\sigma}_\mu z\,,\quad
\dot p_\mu=-\frac{\partial}{\partial z}\stackrel{-}{\sigma}_\mu
\frac{\partial}{\partial\bbar z}
\label{eq64}
\end{equation}
where $\stackrel{\pm}{\sigma}_\mu=(\vec\sigma\,,\pm i)$ are the Pauli
matrices. The formulae of the extended Fock representation will play
the extremely important role in further specific calculations.

\section*{VII. Representation structure of the algebra $h^{(*)}_8$
on the space ${\cal F}_0$ and the fermion-antifermion asymmetry}
      It is meaningful to consider a restriction of the algebra
$h^{(*)}_8$ from space ${\bf F}={\cal F}_F\otimes{\cal F}_0$ on the
subspace of additional variables ${\cal F}_0$ accompanied actually
by restricting of the algebra $h^{(*)}_8$ up to the algebra
$h_4^{(*)}=h_2\oplus h^*_2$. We shall denote generators of the algebra
$h^{(*)}_4$ by $\phi\,,\bar\phi$ which are written down as
\begin{equation}
\phi={{\partial}/{\partial\bbar z}\choose z},\quad
\bar\phi=\left (\bbar z\,,-\frac{\partial}{\partial z}\right)
\label{eq65}
\end{equation}
and act on ${\cal F}_0$. The bilinear forms
$\bar\phi_\alpha\phi_\beta\,,\phi_\alpha\phi_\beta\,,
\bar\phi_\alpha\bar\phi_\beta$ will form the algebra $sp^{(*)}(2,{\bf C})$.
From them real variables written down as
$L_\mu=\bar\phi\sigma_\mu\phi\ (\mu=0,1,2,3)$ will form the algebra
$u(1,1)$. Exlicitly we have ($L_\pm=L_1\pm iL_2$)
\begin{displaymath}
L_+=\bbar zz\,,\quad L_-=-\frac{\partial}{\partial z}
\frac{\partial}{\partial\bbar z}\,,\quad
L_3=-\frac{1}{2}\left (z\frac{\partial}{\partial z}+\bbar z
\frac{\partial}{\partial\bbar z}+1\right),
\end{displaymath}
\begin{displaymath}
L_0=-\frac{1}{2}\left (z\frac{\partial}{\partial z}-\bbar z
\frac{\partial}{\partial\bbar z}+1\right).
\end{displaymath}
Note that $L_+\,,L_-\,,L_0$ is that remains after the restriction
$h^{(*)}_8\supset h^{(*)}_4$ for the operators $p_\mu\,,\dot p_\mu\,,
\frac{1}{2}A$, thus $p_\mu\to (0,0,L_+\,,iL_+)$ and
$\dot p_\mu\to (0,0,L_-\,,-iL_-)$, i.\ e.\ the representation restriction
from ${\bf F}$ to ${\cal F}_0$ is accompanied by restricting of the group
(vector) space $T_{3,1}\supset T_{1,1}$.

      Further, it is easy to see that a representation of the algebra
$u(1,1)$ in the class of functions of two complex variables $z$ and
$\bbar z$ (it will be specified hardly later) has the following structure
\begin{equation}
\left (\bigoplus_{k=0}^\infty D^+(-\frac{_{k + 1}}{^2})\right)
\bigoplus\left (\bigoplus_{k=1}^\infty D^+(\frac{_{k-1}}{^2})\right).
\label{eq66}
\end{equation}
Here $D^+(-\frac{k+1}{2})$ is an irreducible representation of
$su(1,1)$ with a weight $-\frac{k+1}{2}\ (k=0,1,2,\ldots)$,
which is realized in the class of entire functions in the
form $z^k\,f(\bbar zz)$ where
$f(\bbar zz)$ is a entire function of the variable $\bbar zz$, i.\ e.\
of functions expanded in a Taylor series. Vectors of a subspace on
which the representations $D^+(-\frac{k+1}{2})$ are realized
are characterized by positive values of the operator
$F=-2L_0-1=k\geq 0$ standing for the fermionic charge.
In regard to representations $D^+(\frac{k-1}{2})$ with a weight
$\frac{k-1}{2}\ (k=1,2,\ldots)$ which are realized in the class of
meromorphic functions expanded in a Loran series, it
is incompletely reducible and has the structure
\begin{equation}
D^+(\frac{_{k-1}}{^{2}})=D(\frac{_{k-1}}{^2})\ +\!)\
D^{'+}(-\frac{_{k+1}}{^2})
\label{eq67}
\end{equation}
where $D(\frac{k-1}{2})$ is a finite dimensional representation of
$su(1,1)$, and $D^{'+}(-\frac{k+1}{2})$ is its infinite dimensional
``tail'' equivalent to an irreducible representation
$D^+(-\frac{k+1}{2})$, see \cite{2a}. The half-obstinate sum in
(\ref{eq67}) turns in the direction of the invariant subspace.
If the ``tail'' $D^{'+}(-\frac{k+1}{2})$ is realized in the class of
entire functions in the form $\bbar z^k\,f(\bbar zz)$, the finite
dimensional representation is realized in the class of singular (in zero)
functions in the form $\frac{1}{z^k}P(\bbar zz)$ where $P(\bbar zz)$ is a
polinomial of degree $\leq k$ for $\bbar zz$. Vectors, on which
representations $D^+(\frac{k-1}{2})$ are realized, are characterized
by negative values of the fermionic charge $F<0$. It is clear that
singular functions are acceptable neither with the physical nor
mathematical point of view, and therefore it is necessary them to
reject. However to reject only finite dimensional blocks in (\ref{eq66}),
to retain invariant subspaces $D^{'+}(-\frac{k+1}{2})$,
it is impossible. First, $D(\frac{k-1}{2})$ are non-invariant
subspaces, and consequently the factorization on them is not allowable.
Secondly, it is impossible to avoid blocks $D(\frac{k-1}{2})$ (for the
same reason), as only from them it is possible to get in
$D^{'+}(-\frac{k+1}{2})$ with the help of the raising operator $L_+$.
There is only one opportunity to exclude the singular functions ---
it is wholly to exclude (to factorize) all the representations
$D^+(\frac{k-1}{2})$, i.\ e.\ as a representation space of the
algebra $h^{(*)}_4$, to consider the space
\begin{equation}
{\cal F}_0=\bigoplus_{k=0}^\infty{\cal F}_0^{(k)}
\label{eq68}
\end{equation}
where ${\cal F}_0^{(k)}$ is a class of functions in the form
$z^k\,f(\bbar zz)$. On the space the fermionic charge
takes only on positive values. And this means that the
fermion-antifermion symmetry completely broken underlies
the propounded theory. This conclusion is of fundamental
importance for cosmology: at the moment of the creation of our
universe the quantum transition $f\to\dot f$ will be generated
only fermions.

\section*{VIII. Other basic consequences}
      The makings, occurring in space-time, are considered in the
kinetic theory which is classical in its character. In the quantum
theory the makings are not realized in space-time, but in a
representation space, and consequently are essentially probable in
character. If processes proceeding in space-time are evolutionary and
a cause-effect connection is typical for them, quantum processes (in
particular, jumps) follow another principle --- one of
purposefulness, in the case a causality is completely away. From
the pure philosophical point of view, it is necessary to estimate
a making (being of fundamental importance) of the complete symmetry of
the Lorentz group, which
connected with the theorem (\ref{eq63}), and also another so important
making of four-dimensionality of ${\bf R}_{3,1}$ from two-dimensionality of
${\bf R}_{1,1}$, which connected with expansion of a representation of the
algebra $h^{(*)}_8$ from the space ${\cal F}_0$ up to the space ${\bf F}$
(see the previous section). It is impossible to answer a question in
which the moment of time (time is still not present as a category) the
non-Fock representation of the algebra $h^{(*)}_8$ becomes the extended
Fock representation and, hence, when the complete symmetry arises from
the broken Lorentz symmetry in the same way as it is impossible to
answer a question when a representation of $h^{(*)}_8$ extends from
${\cal F}_0$ (connected with two-dimensionality) to ${\bf F}$
(connected with four-dimensionality). All these processes are
essentially probable in character and do not obey the principle of
causality. Thus ``the present'' can be determined by ``the future''.
This happens to be the case for such parameters of
states of relativistic bi-Hamiltonian system as $T_f$ and $T_{\dot f}$
(they will appear in a further consideration), the numerical values
of which are determined by conditions in which the process of quantum
transition $f\to\dot f$ occurs. Before the transition it is
necessary to speak about them as about a priori variables which,
by the way, essentially determines a geometry of space-time continuum
in which there is (before the transition $f\to\dot f$) an ensemble
of states $f$ \cite{1}. In the propounded theory the mentioned above
processes, as if they are pressed all together, exist and go
``simultaneously'', determining the fact that the Leibnitz called
by the beforehand-established harmony (in this case it is
beforehand-established by the algebra $h^{(*)}_8$). Only after
the creation of space-time continuum the processes develop as a
chain of cause-effect events in it.

      One more consequence of the non-unitary theory concerns a
unequivalence of the Heisenberg picture and the Schr$\rm\ddot o$dinger
picture, see \cite{1}, over which Dirac \cite{13} seriously thought
in the framework of the usual unitary scheme, but which really finds out
only within the framework of the non-unitary scheme using the dual pair of
topological vector spaces. By the way, the pair of spaces
$(\Phi'\,,\Phi)$ and the Gauss decomposition (associated with it)
$B_+B_-$ of the group $Sp^{(*)}(4,{\bf C})$ (such that $e^{ipx}\in B_+$,
and $e^{i\dot p\dot x}\in B_-$; in \cite{1} the correspondence
refers to as a polarization of relativistic bi-Hamiltonian system)
result not only in $T$ (and, hence, $CP$)-asymmetry of the
non-unitary scheme (see earlier), but also $P$- and $C$-asymmetry. In
the addition to this we shall notice that $T$-symmetry is broken also
by additional variables $z\dot =z_2$~. Indeed, under the reflection
of $T$, as is known, we have $z_2\to i\bbar z_1$. However
additional variables $z'\dot =z_1$ is not present in the scheme
such that the space ${\cal F}_0$ is not invariant under the
reflection of $T$.

      On the frequently given question why our space has three
dimensions (Kant, Poincar$\rm\acute e$, Erenfest, Weyl) now it is possible
to answer so: three dimensions of space (and its appearence) is caused
by symmetry properties of the algebra $h^{(*)}_8$ and its
representations describing the Universum as a single whole.

      Our discussion will be incomplete, if not to note that
the algebra $h^{(*)}_8$ represents a real
form of the Penroise twistor algebra $h_8({\bf C})$ \cite{14}
which is appeared on the pure geometrical way in the context of the
cohomology theory. For the theory of elementary particles the twistor
program has resulted in rather limited results. However, if to
proceed to the algebra $h^{(*)}_8$ from the point of view of functional
analysis and the theory of representations, it is possible to achieve
an essential progress in construction of the consecutive theory of
elementary particles. These questions will be considered in other articles.


\begin{thebibliography}{96}
\bibitem{1}
S.~S.~Sannikov and I.~I.~Uvarov, Izvestiya Vysshikh Uchebnykh
Zavedenii, seriya Fizika, No.~10, 5(1990)(translation in Soviet
Physics Journal); S.~S.~Sannikov, Ukrain.~J.~Phys. {\bf 40},
No.~7, 650(1995).
\bibitem{2a}
S.~S.~Sannikov, Ukrain.~J.~Phys. {\bf 9}, No.~10, 1139(1964);
\bibitem{3}
J.~von~Neuman {\it Mathematical Foundations of Quantum Mechanics},
translated by R.~T.~Beyer (Princeton University, Princeton, NJ, 1955).
\bibitem{4}
V.~Bargmann, Ann. of Math. {\bf 48}, No.~3, 568(1947).
\bibitem{f1}
The existence of the ground state is connected extremely with the
condition of Hermitian symmetry (\ref{eq9}). Indeed, in the equation
$a^2a^1 f=mf$ there is the number $m=\frac{(a^1f,a^1f)}{(f,f)}\geq 0$
such that $m=0$, if $a^1f=0$. So in this case $m\in{\bf Z_+}$.
\bibitem{2b}
S.~S.~Sannikov, Ukrain.~J.~Phys. {\bf 10}, No.~6, 684(1965);
\bibitem{2c}
S.~S.~Sannikov, Bul.~Inst.~Pol.~Iasi {\bf XII(XVI)},
Fasc. 3-4, 121(1966); Yad.~Fiz. {\bf 1}, No.~9, 570(1965);
Yad.~Fiz. {\bf 6}, No.~6, 1294(1967); Yad.~Fiz. {\bf 4}, No.~3,
587(1966); Nucl.~Phys. {\bf B1}, 594(1967);
\bibitem{2d}
S.~S.~Sannikov, Nucl.~Phys. {\bf B1}, 686(1967);
\bibitem{2e}
S.~S.~Sannikov, preprint ITP-89-45R (Kiev, 1989);
\bibitem{2f}
S.~S.~Sannikov, Problems of nuclear physics and space beams,
No.~32, 31(Kharkov, 1989);
\bibitem{2g}
S.~S.~Sannikov, Teor.~Math.~Phys. {\bf 34}, No.~1, 34(1978);
Ukrain. J. Phys. {\bf 12}, No.~5, 872(1967);
\bibitem{2h}
S.~S.~Sannikov, Dokl.~Akad.~Nauk~SSSR {\bf 209}, 324(1973);
\bibitem{f2}
A weight of the representation $[\lambda_0,c]$ are connected with
$C$ and $C'$ by the formulae $C=\lambda^2_0+c^2-1,C'=i\lambda_0c$,
thus the Pauli pair $(\lambda,\lambda')$ is expressed by
$\lambda,\,c$ under the formulae $\lambda=\frac{1}{2}(\lambda_0+c-1),\,
\lambda'=\frac{1}{2}(c-\lambda_0-1)$.
\bibitem{5}
E.~Majorana, Nuovo Cim. {\bf 9}, 335(1932).
\bibitem{2i}
S.~S.~Sannikov, Dokl.~Akad.~Nauk~SSSR {\bf 176}, 801(1967);
\bibitem{f3}
The algebra $su^c(2)\approx sl(2,{\bf C})$ should not be mixed up
with another algebra $sl(2,{\bf C})$ represented in ${\cal F}_\lambda$
by the operators $(\vec L,\vec N)$, see above.
\bibitem{2j}
S.~S.~Sannikov, Dokl.~Akad.~Nauk~SSSR {\bf 178}, 800(1968);
\bibitem{6}
I.~S.~Granshtein and I.~M.~Ryzhik {\it Tables of Integrals, Series and
Products} 4th ed. (Academic, New York, 1965).
\bibitem{2k}
S.~S.~Sannikov, Dokl.~Akad.~Nauk~SSSR {\bf 177}, 1285(1967);
D.\ Phys.-Math.\ Sc.\ thesis (Institute for Theoretical Physics,
Kiev, 1991);
\bibitem{2l}
S.~S.~Sannikov, Dokl.~Akad.~Nauk SSSR {\bf 198}, 297(1971);
\bibitem{7}
L.~H.~Ryder {\it Quantum Field Theory} (Cambridge University Press,
Cambridge, 1985).
\bibitem{f4}
In particular, the series $\sum^\infty_{n=-\infty}z^n$
summable to the Dirac $\delta (1-z)$-function
has the same property. Note that Euler puts this series equal
to zero by error which now is well clear.
\bibitem{9}
W.~Heisenberg {\it Introduction to the Unified Field Theory of
Elementary Particles} (Intersciences Publishers, London, 1966).
\bibitem{10}
M.~A.~Naimark {\it Linear representations of the Lorentz group}
(Pergamon, Oxford, 1964).
\bibitem{11}
L.~de~Broglie, Rev. d'histoire des science {\bf 5}, 289(1952).
\bibitem{2m}
S.~S.~Sannikov, preprint ITP-67-34 (Kiev, 1967);
\bibitem{12}
P.~A.~M.~Dirac, Proc. Royal Soc. {\bf D10}, No.~6, 1760(1974).
\bibitem{f5}
The appearence of a double sum in (\ref{eq53}) is a simple
consequence of that $su^c(2)$ is a Lie algebra of the rank 2 in which
a representation are characterized by a pair of independent numbers
$(\lambda\,,\lambda')$ satisfying only one condition
$\lambda-\lambda'\neq\frac{n}{2}$.
\bibitem{2n}
S.~S.~Sannikov, Ukrain.~J.~Phys. {\bf 11}, 106(1966);
\bibitem{2o}
S.~S.~Sannikov, preprint ITP-73-6R (Kiev, 1973);
\bibitem{13}
P.~A.~M.~Dirac {\it Lectures on Quantum Field Theory}
(Yeshiva University, New York, 1967).
\bibitem{14}
R.~Penrose, Rep. on Math. Phys. {\bf 12}, 65(1977).
\end{thebibliography}
\end{document}